\documentclass[useAMS, usenatbib]{mn2e}
\usepackage{aas_macros}
\usepackage{amsmath}
\usepackage{amssymb} 
\usepackage{graphicx}
\usepackage{subcaption} 
\usepackage{float}
\usepackage{booktabs} 
\usepackage{multirow} 
\usepackage{amsfonts}
\usepackage{lmodern}
\usepackage{bm}
\usepackage{upgreek}

\newcommand{\bc}{\begin{center}}
\newcommand{\ec}{\end{center}}

\renewcommand{\vec}[1]{\mathbf{#1}}
\let\oldhat\hat
\renewcommand{\hat}[1]{\oldhat{\mathbf{#1}}}
\let\oldbullet\bullet \renewcommand{\bullet}[1][0pt]{%
\mathrel{\raisebox{#1}{$\oldbullet$}}%
}

\title{Resolving flows around black holes: numerical technique and
  applications} \author[Curtis \& Sijacki]{Michael Curtis$^1$\footnotemark[1], Debora
  Sijacki$^{1}$ \\
  $^1$ Institute of Astronomy and Kavli Institute for Cosmology,
  University of Cambridge, Madingley Road, Cambridge CB$3$ $0$HA, UK}

\begin{document}

\maketitle

\begin{abstract}
Black holes are believed to be one of the key ingredients of galaxy formation
models, but it has been notoriously challenging to simulate them due to the
very complex physics and large dynamical range of spatial scales
involved. Here we address a significant shortcoming of a
Bondi-Hoyle-like prescription commonly invoked to estimate black hole
accretion in cosmological hydrodynamic simulations of galaxy formation,
  namely that the Bondi-Hoyle radius is frequently unresolved. We describe
and implement a novel super-Lagrangian refinement scheme to increase,
adaptively and 'on the fly', the mass and spatial resolution in targeted
regions around the accreting black holes at limited computational cost. While
our refinement scheme is generically applicable and flexible, for the purpose
of this paper we select the smallest resolvable scales to match black holes'
instantaneous Bondi radii, thus  effectively resolving Bondi-Hoyle-like
accretion in full galaxy formation simulations. This permits us to not only
estimate gas properties close to the Bondi radius much more accurately, but
also allows us to improve black hole accretion and feedback
implementations. We thus devise a more generic feedback model where accretion
and feedback depend on the geometry of the local gas distribution and where
mass, energy and momentum loading are followed simultaneously. We present a
series of tests of our refinement and feedback methods and apply them to
models of isolated disc galaxies. Our simulations  demonstrate that resolving
gas properties in the vicinity of black holes is necessary to follow black
hole accretion and feedback with a higher level of realism and that doing so
allows us to incorporate important physical processes so far neglected in
cosmological simulations. 
\end{abstract}


\section{Introduction}
\renewcommand{\thefootnote}{\fnsymbol{footnote}}
\footnotetext[1]{E-mail: mc636@ast.cam.ac.uk}

Growing observational evidence indicates that supermassive black holes reside
in the centres of the majority of galaxies, including our own Milky
Way. Observed scaling relations between the mass of black holes and the
properties of the host galaxy bulge such as its mass, luminosity and velocity
dispersion \citep[e.g.][]{Kormendy:95, Magorrian:98, Ferrarese:00, Gebhardt:00,
  Marconi:03, Gultekin:09, Kormendy:13, McConnell:13} suggest
that the evolution of the central black holes and the host galaxies are
closely linked. However, the extent to which the formation and growth of black
holes impacts the galaxy formation process is not fully understood.

More direct evidence for the effects of black holes comes from recent
observations of active galactic nuclei (AGN)-driven winds
\citep[e.g.][]{Pounds:03, Sturm:11, Maiolino:12, Veilleux:13, Cicone:14,
  Genzel:14, Tombesi:15}, indicating that AGN activity is affecting host
galaxies by, for  example, expelling large amounts of material at thousands of
$\mathrm{km} \, \mathrm{s}^{-1}$. Moreover, in massive galaxies residing in
group and cluster environments, a large body of X-ray studies \citep[see
  e.g.][for two well studied examples]{Fabian:06, Forman:07} indicates that
black hole-launched jets directly interact  with the surrounding medium,
heating it and thus suppressing star formation in the central galaxy. The
umbrella term  for these processes is `feedback' \citep[for a recent review
  see][]{Fabian:12}, and this is understood to result from accretion of matter
on to the black hole and the subsequent release of energy
\citep[][]{LyndenBell:69}. 

Since the mutual interaction between black holes and their host galaxies
became apparent, it has become necessary to understand the accretion processes
involved in order to answer many open questions in the field, including where
the black hole--host relations come from, and the role AGN feedback plays in
the quenching of star formation and in the evolution of galaxy
morphologies. Various models have been proposed to describe the physics of
black hole accretion  and feedback analytically \citep[e.g.][]{Silk:98,
  Fabian:99b, King:05}. Such models have generally been successful in
explaining the black hole mass -- stellar bulge velocity relationship (the
so-called $M_\mathrm{BH}-\sigma$ relationship). They are, however, inevitably
forced into making simplifying assumptions, such as spherical geometry, lack
of treatment of concurrent star formation and large scale cosmological inflows
of gas, that might seriously hinder their predictive power \citep[see recent
  work by][]{Costa:14}. It is worth noting that accretion is not only coupled
to the gas hydrodynamics, which is highly non-linear and complex, but also to
the gravitational torques occurring on a large range of scales which are also
difficult to describe analytically.

There is clear motivation, then, for including an accurate treatment of black
hole growth and feedback in cosmological hydrodynamic simulations of galaxy
formation and there have been many studies to this effect over the past 10
years \citep[e.g.][]{DiMatteo:05, Kawata:05, Springel:05, Sijacki:07,
  Booth:09, Debuhr:11, Power:11, Choi:12, Dubois:12}. These models have to
overcome the numerous computational and theoretical challenges associated with
simulating AGN physics. First, and perhaps most importantly, the spatial
scales involved span a massive range. If even a basic statistical sample of
galaxies is to be achieved then a simulation domain must cover tens to
hundreds of mega-parsecs (Mpc). This should be contrasted with the
Schwarzschild radius for a supermassive black hole - for a mass of $10^7
M_{\odot}$, $r_\mathrm{s} \approx 10^{-6}\, {\rm pc}$: a dynamic range that spans 14
orders of magnitude.

Thus if we want to study the impact of AGN on galaxies in full cosmological
volumes, directly resolving the region of accretion is computationally
infeasible and will continue to be so for a number of years. Furthermore, even
if we could achieve the resolution necessary, the physics that governs
accretion is poorly understood. To overcome this, numerical simulations that
include AGN physics must adopt so-called sub-grid models. These algorithms
track the unresolved physical processes that are ultimately expected to impact
the simulations on large scales. In the case of AGN, these prescriptions
define the changes in the physical properties of the black holes within the
simulation, including how they accrete mass from the surrounding gas, how they
are advected with the fluid flow, and how they deposit mass, energy and
momentum back into the surrounding medium.

Many sub-grid techniques have been proposed and implemented in previous works
\citep[e.g. see a recent comparison study by][]{Wurster:13}, but there are
essentially two main strategies. The first is to allow black holes to swallow
particles or cells that come within a certain distance of them. Whilst this
has the benefit that it does not require any further approximation, it has the
downside that it is highly stochastic, causing the black hole mass to grow in
discontinuous jumps as particles randomly fly close enough to be
swallowed. Furthermore, there is no reason to expect such an accretion rate to
be reliable at the typical resolution of current simulations. 

The second approach to quantifying the gas accretion is to parameterize an
accretion rate in terms of the local gas properties. Such a rate may be
theoretically motivated, or may be a parameterization from higher resolution
`zoom-in' simulations of smaller spatial scales. Perhaps the most commonly
used theoretical rate is a Bondi-Hoyle-Lyttleton approach \citep{Bondi}, which
adopts the simple formula
\begin{equation}
\dot{M}_\mathrm{B} = \frac{4\uppi G^2 M^2_\mathrm{BH} \rho_\infty}{(c^2_\infty + v^2)^{3/2}},
\end{equation}
where $M_\mathrm{BH}$ is the mass of the black hole, $\rho_\infty$ and
$c_\infty$ are the gas density and sound speed at infinity and $v$ is the
relative velocity between the black hole and the gas at infinity. Whilst
elegant and easy to implement numerically, this approach also has 
a number of drawbacks. 

First, for the Bondi calculation to be accurate, we require that the fluid
be resolved at the Bondi-Hoyle radius\footnote{Note that in the absence of the
  relative velocity term this simplifies to the Bondi radius $r_\mathrm{B} =
  \frac{GM_\mathrm{BH}}{c^2_\infty}$, which we will use throughout the paper.}
\begin{equation}\label{BR}
r_\mathrm{B} = \frac{GM_\mathrm{BH}}{c^2_\infty + v^2},
\end{equation}
which is not the case, either for the vast majority of cosmological
simulations, or for idealized test problems in the past literature.

 Secondly, the Bondi rate ignores the rotation of the gas, which is likely
  to be very important. Gas with enough angular momentum will circularize
  before reaching the inner most stable orbit, forming an accretion disc and
  thus possibly slowing the rate at which such matter can fall on to the black
  hole. There have been several studies that have studied gas angular momentum
  near supermassive black holes in galaxy formation simulations \citep[see
    e.g.][]{Levine2010, Hopkins2011} or that have directly attempted to
  account for the angular momentum within an accretion rate prescription
  \citep[e.g.][]{Krumholz, Power:11, A-A:13, Rosas:14}. These attempts face a
  number of key challenges: gas orbits must be accurately followed to
  robustly measure the angular momentum of the accretion flow on resolvable
  scales, whilst assumptions about the properties of the (innermost)
  accretion disc need to be made - in particular, the nature and magnitude of
  the viscous processes that drive the evolution of such discs. The gravitational
  instability of accretion discs leading to fragmentation
  \citep{KingPringle2007} and possibly star formation and stellar feedback add
  yet another non-trivial layer of  complexity.

 Furthermore, the Bondi-Hoyle rate ignores several other effects, such as
  the impact of a non-uniform, turbulent gas flow \citep{Krumholz:06}. In this
  case, the deviation from the classic formula is less obvious, as for higher
  Mach number flows the accretion rate may be significantly increased, but
  such an enhancement can be offset by the vorticity of the
  flow. Additionally, in a regime where cooling is efficient, gas may be in
  free fall at large distances from the black hole leading to a significantly
  different outcome with respect to the standard Bondi-Hoyle rate
  \citep{Hobbs:12}. 

  Despite these limitations, the Bondi-Hoyle approach is still widely adopted
  in cosmological simulations. The extent to which the simulated accretion
  on to the black holes may deviate from reality is, so far, unclear, since
  the other key element in our models of black hole growth - the injection and
  coupling of feedback to the surrounding gas - is also poorly
  constrained. Understanding both the feedback and the accretion together is
  fundamental - if the feedback is sufficient to shut off accretion then the
  black holes may enter a period of self-regulation, rendering the precise
  details of the accretion rate less important.

The goal of this paper is to tackle the first of these shortcomings, the
  unresolved Bondi-Hoyle radius, which is inherent in the current use of
Bondi accretion in cosmological simulations. To this end we present a new
modelling technique that enables us to resolve, 'on-the-fly' and adaptively,
the gas surrounding black hole particles in a super-Lagrangian fashion. We use
this technique to improve our estimation of the fluid properties local to the
black hole, which we then use to compute the Bondi-Hoyle-like accretion more
accurately. Our modelling technique has the potential to tackle
  additional shortcomings of the Bondi-Hoyle model, such as the need to
  consider the angular momentum barrier, or to study accretion on much smaller
  scales, but these will be left for future work. Thus in this work we still rely on a simple Bondi-like sub-grid model for gas accretion which, as detailed above, has significant limiting assumptions.

In addition to our novel refinement scheme, we implement concomitant changes
to the algorithm by which black 
hole particles accrete and are advected, as well as the process by which mass
is drained from the surrounding gas cells. The large gain in the mass and
spatial resolution around black holes allows us to implement a more
comprehensive black hole feedback model where mass, energy and momentum are
imparted to the surrounding medium and where the resolved gas geometry
dictates feedback geometry. We validate our implementation with a suite of
simulation tests, before investigating the effect of these changes on models
of isolated galaxies. While our refinement method allows us to track gas
angular momentum much more accurately, for the purposes of this work we
neglect the impact of gas angular momentum on the Bondi rate, which will be
the main topic of our follow up paper. In addition, any such model will
  always be affected and potentially limited by its treatment of star
  formation and the structure of the interstellar medium (ISM). Ultimately to
  improve the accuracy 
  of the modelling of black hole accretion it is clear that the subgrid models
  that govern star formation will also need to be improved. 

The paper summary is as follows. In Section~\ref{Methodology} we present our
methodology and explain the details of our new implementation. In
Section~\ref{Results:Bondi} we validate our implementation and choice of
parameters on classical Bondi inflow solutions, while in
Section~\ref{Results:Galaxy} we present the results of our new refinement
technique and black hole feedback on models of isolated, disc-dominated
galaxies. Finally, in Section~\ref{Conclusions} we present our conclusions and
discuss future work.  

\section{Methodology}
\label{Methodology}
\subsection{Code}
\subsubsection{Basic Setup}

All simulations in this paper are carried out using the {\small AREPO} code
\citep{Springel:10} which employs a moving mesh based on a quasi-Lagrangian
finite volume technique. {\small AREPO} handles gravitational interactions
using the TreePM approach \citep{Springel:05}, and dark matter is modelled by
a collisionless fluid of massive particles. In contrast to traditional
adaptive mesh refinement (AMR) codes, which usually employ a Cartesian grid of
cells that are then refined and de-refined according to some criteria, {\small
  AREPO} generates an unstructured mesh based on the Voronoi tessellation of a
set of discrete points that cover the computational domain. These
mesh-generating points are allowed to move freely with the local velocity of
the flow, with some subdominant corrections to allow for mesh
regularisation. In this way, {\small AREPO} retains many of the advantages of
smoothed particle hydrodynamics (SPH) techniques such as Galilean invariance
and  resolution that naturally and continuously follows the fluid flow, as
well as advantages of AMR methods such as better resolution of shocks, contact
discontinuities and fluid instabilities \citep[see  e.g.][]{Bauer:12, Keres:12, Sijacki:12, Torrey:12,
  Vogelsberger:12}. In particular, for the
purpose of this work, we exploit the code's ability to adaptively refine or
de-refine the Voronoi mesh based on flexible criteria. Note that
  particle splitting techniques have been employed to allow for adaptive mass
  resolution in SPH codes \citep[e.g.][]{Kitsionas:02, Martel:06, Mayer:15},
  while in AMR codes cells can be refined based on a number of predefined
  criteria, such as quasi- or super-Lagrangian refinements
  \citep[e.g.][]{Teyssier:02, Chapon2013}.  Refinement in  {\small AREPO}
  allows for smooth variations between regions of different mass resolution
  without imposing (relatively) sharp boundaries that may be prone to
  diffusion errors and numerical heating. The ability to de-refine in {\small
    AREPO} (that is, remove regions of high resolution) using a natural
  merging technique, is particularly important since refined particles may be
  moved large distances from a black hole by expulsive feedback.

For a subset of our simulations we adopt primordial gas cooling and a star
formation sub-grid model as in \citet{SF}. Gas cells above a density threshold
of $\rho_\mathrm{sfr} = 0.18\, \mathrm{cm}^{-3}$ stochastically form star
particles, with a maximum formation time of $t_\mathrm{sfr} = 1.5\, {\rm
  Gyrs}$, which then only interact gravitationally with other particles in the
simulation. Our model assumes that the ISM is in a
self-regulated equilibrium state, representing the cumulative effect of
unresolved thermal and turbulent processes, and that as such we can relate the
average temperature as a function of density using an effective equation of
state. To study the effect of colder gas present in the vicinity of black
holes, we include metal line cooling in some simulations. Where present, we
calculate this using the routine outlined in \citet{Vogelsberger:13} whereby
cooling rates are calculated based upon the rates for a Solar abundance gas,
scaled linearly with the total metallicity. As we are more interested in the
range of the effect that the presence of cold gas can achieve we set the
metallicity of all cells to be that of Solar composition gas. Note that
  as in \citet{Vogelsberger:13} metal line cooling is applied to the gas which
is not in the multi-phase ISM.  
  
\subsubsection{Refinement}

{\small AREPO} applies well tested routines to regularize and (de-)refine the
cells of the Voronoi mesh as appropriate. This results in the code being able
to maintain the cells at a similar target mass, as well as dynamically adapt
to regions of higher density. Cells can also be flagged for refinement based
on nearly arbitrary criteria, and this allows us to increase the spatial
resolution over localized regions of the computational domain. Each cell
selected is then split into two cells by the introduction of an additional
mesh-generating point that is inserted at almost exactly the same location as
that of the original cell, with conserved quantities, namely mass, energy
  and momentum, split between the new cells in a conservative
manner \footnote{ It is worth pointing out that the angular momentum
    conservation is not guaranteed in {\small AREPO}, but as shown in
    \citet{Pakmor2015} in the case of galaxy formation simulations it does not
    lead to any significant errors with respect to the schemes that exhibit
    second-order convergence in angular momentum conservation.}. The
mesh-regularization techniques, which add a small additional corrective term
to the velocity, in the direction of the centre of mass of the cell, mean that
the two points become separated from each other over the course of a few
timesteps. It is worth emphasizing that this is a key difference between
{\small AREPO} and standard AMR codes: there is only ever a single mesh
covering the simulation domain. In {\small AREPO}, by refinement, we mean the
introduction of additional cells to increase the resolution over particular
regions.

\subsection{Novel refinement scheme around black holes}

\begin{figure*}
\centering
\includegraphics[width=0.45\textwidth]{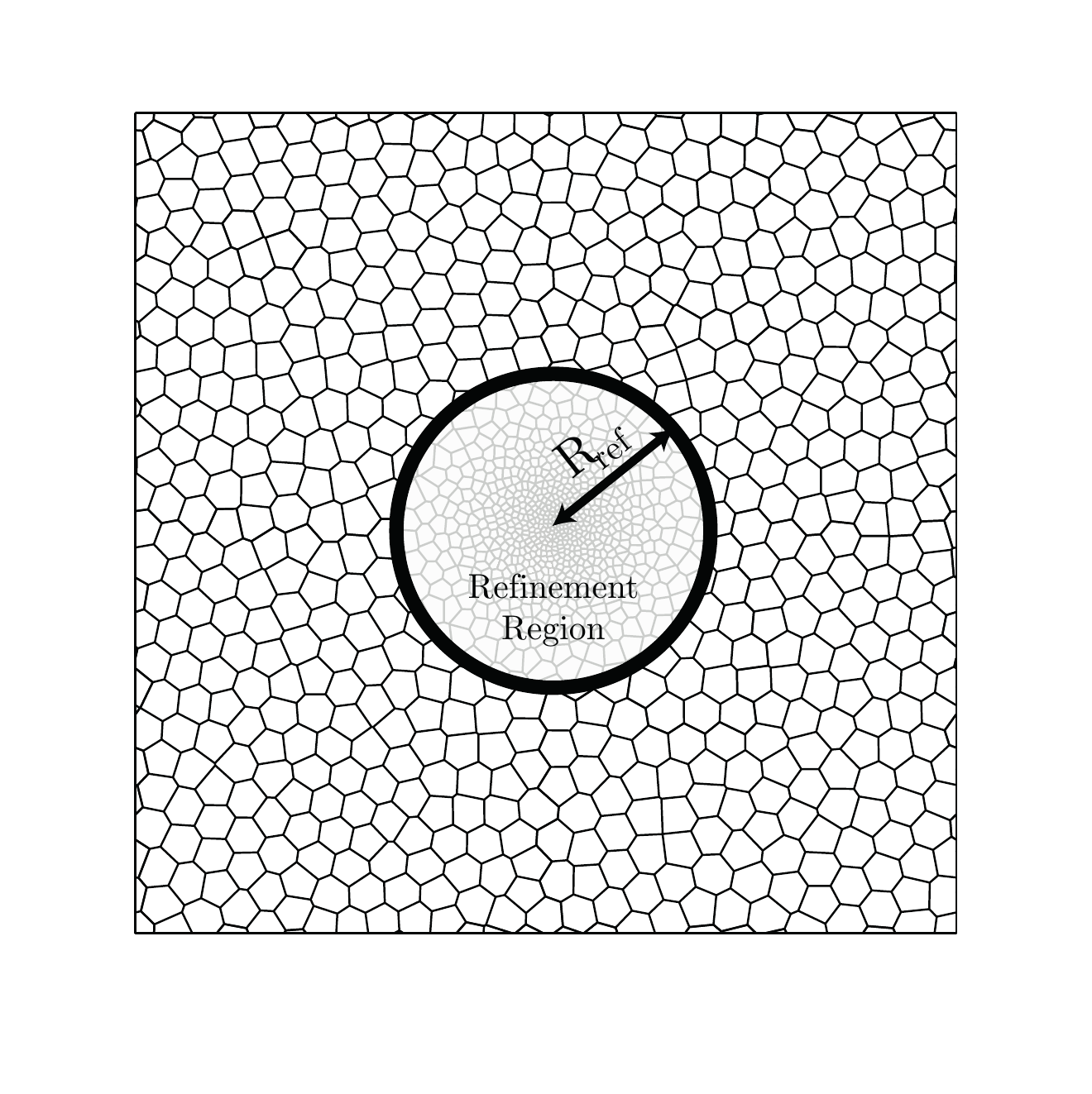}
\includegraphics[width=0.45\textwidth]{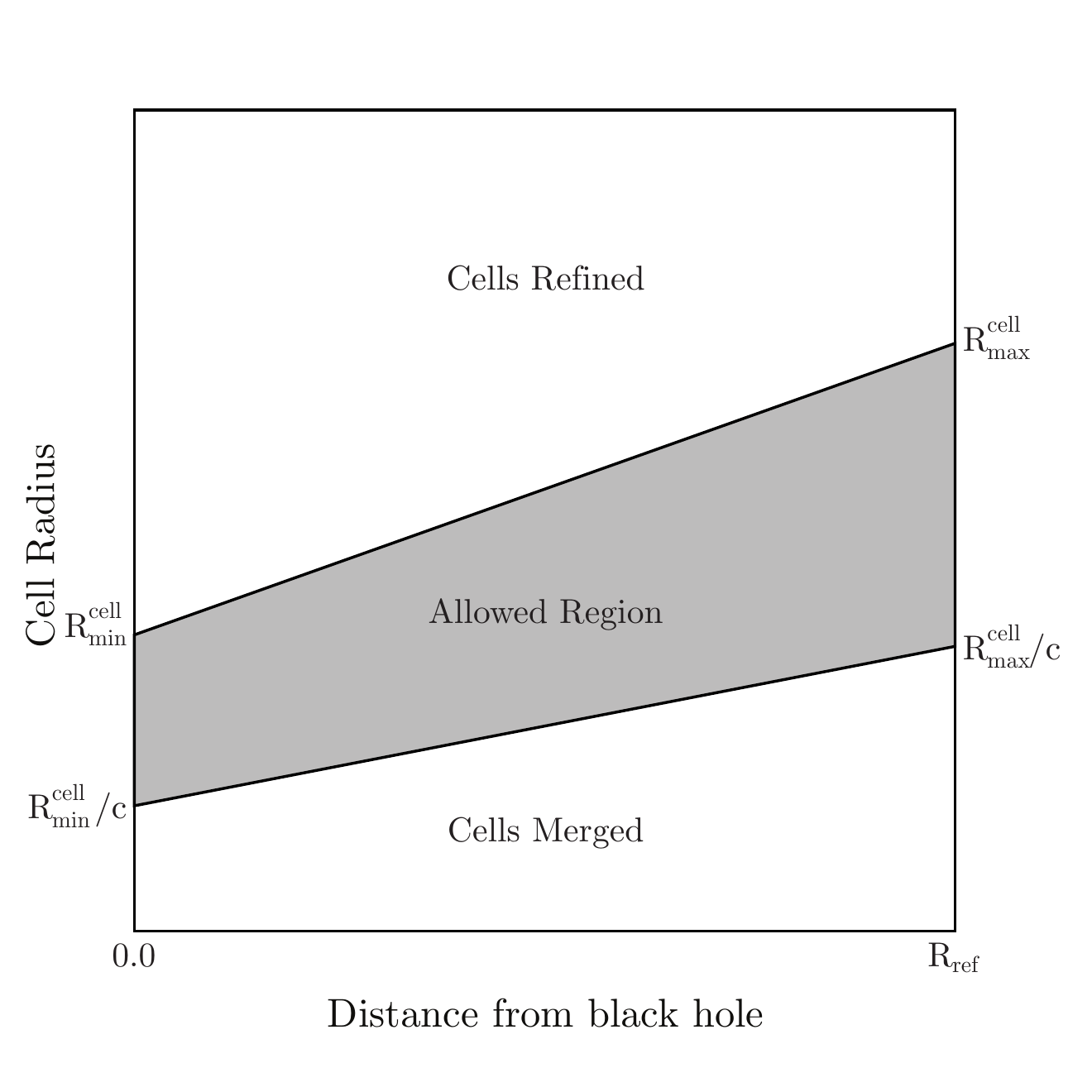}
\caption{During normal code operation, cells are (de-)refined to keep their
    mass close to a pre-specified target value. Using our refinement scheme,
    however, we force cells within adaptively defined refinement radius,
    $R_\mathrm{ref}$, (for a schematic representation, see the left-hand panel) to have a
    radius within the shaded zone indicated in the right-hand panel. If the cells lie
    outside of this shaded region, they are either refined (split into two) or
    de-refined (merged) as appropriate.}  
\label{ref_scheme}
\end{figure*}

As described above, the criterion by which a cell is flagged for refinement
can be almost arbitrary. We use this to our advantage when considering the
problem at hand: improving the resolution and thus the estimation of the local
properties of the gaseous fluid around black hole particles in our
simulations. Here, we adopt a new strategy, by which cells are forced to have
a radius that lies between two values, which increase linearly with distance
from the nearest black hole.

Gas cells with cell radius $R$ that are at a distance of $d$ from a black hole
are flagged for (de-)refinement if $d < R_\mathrm{ref}$ and if they lie
outside of the range (see Figure~\ref{ref_scheme})
\begin{align}
\frac{d}{R_\mathrm{ref}}\frac{(R^\mathrm{cell}_\mathrm{max} - R^\mathrm{cell}_\mathrm{min})}{c} +  \frac{R^\mathrm{cell}_\mathrm{min}}{c} &< R, \\
\frac{d}{R_\mathrm{ref}}(R^\mathrm{cell}_\mathrm{max} - R^\mathrm{cell}_\mathrm{min}) +  R^\mathrm{cell}_\mathrm{min} &> R\;,
\end{align}
where $c$ is a constant parameter that we set to $2$. Our numerical
experiments indicate that this value of $c$ represents a balance between
setting $c$ too high, which would limit refinement or incur a large
computational cost due to an overabundance of smaller cells, and setting $c$
too low, which would mean that cells were forced to lie in a very strict
range, leading to large computational cost as cells are constantly being
refined and de-refined. In addition to 
this, we explicitly suppress stars from forming in the region $d <
R_\mathrm{ref}$. Allowing star particles to form from gas cells with a large
mass spectrum increases the probability of collisions that can cause an
unphysical increase in the kinetic or thermal energy of the smaller
particle. By preventing our refined cells from forming stars we ensure there
are no spurious \textit{N}-body heating effects, which can affect the temperature of
the gas 
local to the black hole, although there could still be some heating caused by dark matter particles moving through the central region of the galaxy. As we will see, the suppression does not have a significant
impact on the overall star formation rate of the galaxy. In future work, we
will develop more sophisticated representations of the ISM to better model
star formation in the vicinity of black holes in order to make full use of our increased resolving power, but this is beyond the scope of
this paper.   

\subsubsection{Choice of parameters}

Within our refinement scheme, there are essentially three parameters to be
defined: the volume over which we should refine which is set by
$R_\mathrm{ref}$ as we assume spherical refinement regions, and the maximum
$R^\mathrm{cell}_\mathrm{max}$ and minimum $R^\mathrm{cell}_\mathrm{min}$
radii, which define the spatial scale at which cells should be refined. Our
strategy for choosing these parameters is motivated by several factors, namely 
\begin{enumerate}
  \item the refinement region should always contain at least a certain minimum mass of gas
  \item the accretion rate should be sufficiently resolved such that it
    converges to the results of higher resolution simulations 
  \item the refinement scheme should not, inadvertently, de-refine cells as
    they enter the refinement radius 
  \item computational overhead should be kept to a minimum
  \item the Bondi radius is the physical scale that we want to resolve.
\end{enumerate}
To guarantee the first of these, it makes sense to set $R_\mathrm{ref}$ to be
a function of the black hole smoothing length, $h_\mathrm{BH}$, which is
defined as the radius of the sphere in which $N_{\rm ngb}\times m_{\rm target}$ is
present. Here $N_{\rm ngb}$ is the number of neighbouring gas cells which we set
to at least $32$\footnote{Note that by default we increase $N_{\rm ngb}$ with
resolution so as to ensure that the total enclosed mass within
$h_\mathrm{BH}$ is constant.}. $m_\mathrm{target}$ is the gas cell mass that
{\small AREPO} maintains outside of the refinement region and which is
typically equal to the gas cell mass at the beginning of the simulation. While
values 
$R_\mathrm{ref}>h_\mathrm{BH}$ give improved resolution over a larger volume,
we find that this choice leads to no significant improvement in the accretion
rate estimate over our simulations with $R_\mathrm{ref}=h_\mathrm{BH}$. This
is because $h_\mathrm{BH}$  by
definition contains a consistent mass which  
we find sufficient for the refinement scheme to be effective. Thus, all
simulations in this paper use $R_\mathrm{ref}=h_\mathrm{BH}$, except where we
explicitly state otherwise. It is worth pointing out that with our black
  hole refinement scheme we could in principle significantly reduce the black
  hole smoothing length (while keeping the refinement region larger) as the
  gas flow is much better resolved. However, for the purposes of this paper we
keep the black hole smoothing length the same as in our non-refined
simulations to allow for a more straightforward comparison between the two.

We find that by also setting $R^\mathrm{cell}_\mathrm{max}$ to be a function
of $h_\mathrm{BH}$ we are able to prevent the refinement scheme from merging
cells inappropriately. After testing a range of different values we find that
$R^\mathrm{cell}_\mathrm{max}=0.5\;h_\mathrm{BH}$ represents a good choice. We
demonstrate this in Section~\ref{Results:Galaxy}, where we study the
distribution of gas cell masses as a function of refinement parameters for
isolated galaxy disc simulations. 

Finally, our goal of resolving scales of the order of the Bondi radius around
the black hole motivates setting $R_\mathrm{cell}^\mathrm{min}$ to be equal to
$r_{\rm B}$ evaluated at each timestep when the black hole particle is
  active. Further to this, we impose a minimum cell mass to avoid excessive
computational cost. The size of this is set by the maximum cell mass needed to
resolve the Bondi radius. This is obviously dependent on the density of the
gas close to the black hole, and so must be set experimentally. We find that a
mass of $m_\mathrm{min} = 10^{-2}M_{\mathrm \odot} \approx
10^{-7}m_\mathrm{target}$, valid for our specific simulations of isolated galaxies, is
sufficient.

\subsection{Black Hole Model}

\subsubsection{Black Hole Mass}
Following previous studies \citep{DiMatteo:05, Springel:05}, we represent
black holes using collisionless, massive sink particles. Our simulations
maintain two separate black hole masses: the `sub-grid' mass
($M_\mathrm{BH,sub}$) and the `dynamical' mass ($M_\mathrm{BH,dyn}$). The
sub-grid mass is the mass of the black hole for the purposes of the feedback
algorithm: at each time step this mass increases by
$\dot{M}_\mathrm{BH}\mathrm{d}t$ and the resulting mass is used to calculate
the subsequent accretion rate. The dynamical mass of the black hole particle
is used in the calculation of the gravitational potential. This mass only
increases via accreting mass from surrounding gas cells or by merging with
other black hole particles.

The need for two masses is driven by the problems of resolving the black hole
properly. Gas cells surrounding the black hole frequently have masses that are
the same order of magnitude as, or even larger than, the black hole
itself. Allowing the black hole to accrete neighbouring gas cells directly
would therefore be unphysical. Ideally, the increase in the sub-grid mass
$M_\mathrm{BH,sub}$ should be paralleled by increases in the dynamical mass
$M_\mathrm{BH,dyn}$ so that the code maintains $M_\mathrm{BH,dyn} =
M_\mathrm{BH,sub}$ as closely as possible, which in practice is true once the
sub-grid mass exceeds the gas cell mass.

The possibly large masses of the gas cells in our simulations relative to the
black hole particle mass (especially in the case of the initial black hole
seed) mean that the black hole is prone to being ejected by unphysical
two-body encounters. We note that this is a problem that is present in all
simulations with comparable dark matter/gas mass and black hole seed
mass. Other studies have forced the black hole position to track the potential
minimum of their host halo. We find, however, that once we resolve the flow
around the black hole this leads to unacceptable movement of the black
hole. In this paper, we instead allow the black hole particle to have a
relatively large dynamical mass. This mass is set to be large enough such that
the black hole is subject to sufficient dynamical friction that it remains in
the centre, but small enough such that we do not introduce large changes to
the central potential. We find that $M_\mathrm{BH,dyn} = 200 \times
M_\mathrm{DM}$ is the sufficient mass required to maintain the black holes
particles in the correct position. Note also that, due to our
super-Lagrangian refinement, the typical gas cell mass in the vicinity of black
holes is much smaller than is the case without the refinement, which further
minimizes artificial black hole displacements.

\subsubsection{Accretion Rate} \label{sec:hsml}
Following \citet{Springel:05}, we take as our fiducial model a
Bondi-Hoyle-like prescription to estimate the accretion on to the black hole 
\begin{equation}
\dot{M}_\mathrm{B} = \frac{4\uppi \alpha G^2 M^2_\mathrm{BH} \rho_\infty}{c_\infty^{3}},
\label{bondi_rate_eq}
\end{equation}
where $M_\mathrm{BH}$ is the mass of the black hole, $\rho_\infty$ is the gas
density at infinity and $c_\infty$ is the sound speed at infinity. The
traditional Bondi-Hoyle-Lyttleton rate also includes the relative velocity of
the black hole. For simplicity, we do not include this here. Here, $\alpha =
100$ is a dimensionless parameter that we use to account for the unresolved 
cold and hot ISM phases \citep[for further discussion
  see][]{Sijacki:11}. In addition, in all simulations we limit the accretion
rate to the Eddington limit 
\begin{equation}
\dot{M}_\mathrm{Edd} \equiv \frac{4\uppi GM_\mathrm{BH}m_\mathrm{p}}{\epsilon_\mathrm{r}\sigma_\mathrm{T}c}\;,
\end{equation}
where $\sigma_{\rm T}$ is the Thomson cross-section, $m_\mathrm{p}$ is the mass of a proton
and $\epsilon_\mathrm{r} = 0.1$ is the radiative efficiency. For
non-spherically symmetric accretion, it is possible for accretion to exceed
the Eddington limit, for example, if energy is transported away by a
collimated outflow \citep{Kurosawa:09, Jiang:14}. The extent to which AGN accrete in super-Eddington regimes is as yet
unclear, but it is not something that we consider in this paper. 

In calculating the fluid properties in the locality of the black hole, we must
average over the central volume weighted by the mass of the gas cells and
(optionally) by some kernel function. In SPH codes this is handled naturally,
as the value of each field centred on the black hole is by default just the
SPH kernel weighted sum evaluated at the black hole's position. Here, we adopt a
similar technique with top-hat kernel and simply calculate the fluid
properties as a mass weighted sum over the black hole smoothing length. We
choose this top-hat approach over other alternatives (such as a cubic spline kernel) as we do not wish to overweight the innermost gas cells in our
calculations given that we are resolving the physical scales relevant for Bondi accretion.

\subsubsection{Cell Draining}

When the sub-grid mass increases, we must drain mass from surrounding gas
cells to ensure mass conservation. To accomplish this we adopt a model similar
to that of the previous simulations of \citet{Vogelsberger:13}, in which up to 90\%
of gas is drained from the primary cell of the black hole particle. Here,
however, our refinement scheme forces cells to become progressively smaller
the closer they are to the black hole. Limiting this process to a single cell
means that there is often not enough mass available - the black hole cannot
keep up and gas builds up causing the accretion rate to rise. We avoid this
problem by allowing mass to be drained from multiple cells surrounding the
black hole. We do this by allowing the sink particle to drain up to 90\% of a
cell's mass in a manner weighted by the mass of the cell. At each time step,
the black hole attempts to drain $\Delta M = (1-\epsilon_\mathrm{r})\dot{M}_\mathrm{BH}
\Delta t$ from the cells within its smoothing length. A cell of mass $m$ is
drained of $\Delta m = \Delta M \frac{m}{m_\mathrm{tot}}$, where
$m_\mathrm{tot}$ is the total mass of gas cells within the black hole
smoothing length.

\subsubsection{Feedback}

Here we adopt two different mechanisms of injecting feedback. In each, the AGN luminosity
is set to be a fraction $\epsilon_\mathrm{r}$ of the rest mass energy
available from accreting material. 

In the first case, feedback is implemented by coupling a fraction
$\epsilon_\mathrm{f}$ of the luminosity thermally and isotropically to the
surrounding gas, so that
\begin{equation}
\Delta E_\mathrm{feed} = \epsilon_\mathrm{f}\epsilon_\mathrm{r}\Delta M_\mathrm{BH} c^2 ;
\end{equation}
$\epsilon_\mathrm{f} = 0.05$ here represents the feedback efficiency, which we
set to reproduce the normalization of the locally inferred $M_\mathrm{BH} -
\sigma$ relation \citep{DiMatteo:05, Sijacki:07}. The energy is distributed
over all cells within the smoothing length of the black hole, weighted
according to  a simple cubic spline kernel, as given by \citet{Monaghan:85}. Note that this is different to the
top-hat kernel that we use for estimation of the fluid parameters for the
Bondi rate. There is no particular reason why these two kernel weightings
should be the same in grid based codes. In this case, we are motivated by
wanting to model energy released in the immediate vicinity of the black hole,
in contrast to the estimation of fluid parameters at a larger distance from
the black hole (i.e. $\rho_\infty$, $c_\infty$) as with the accretion
rate calculation.

In the second case, we inject a total momentum of $L/c$ into the cells
surrounding the black hole. In the simplest case we inject the momentum
isotropically as well \citep[see also][]{Costa:14}. Similar to
above, each cell within the smoothing length of the black hole receives a kick
to its momentum, weighted according to the cubic spline kernel, in the
radial direction away from the black hole. 

\begin{figure*}
\centering
\includegraphics[width=0.3\textwidth]{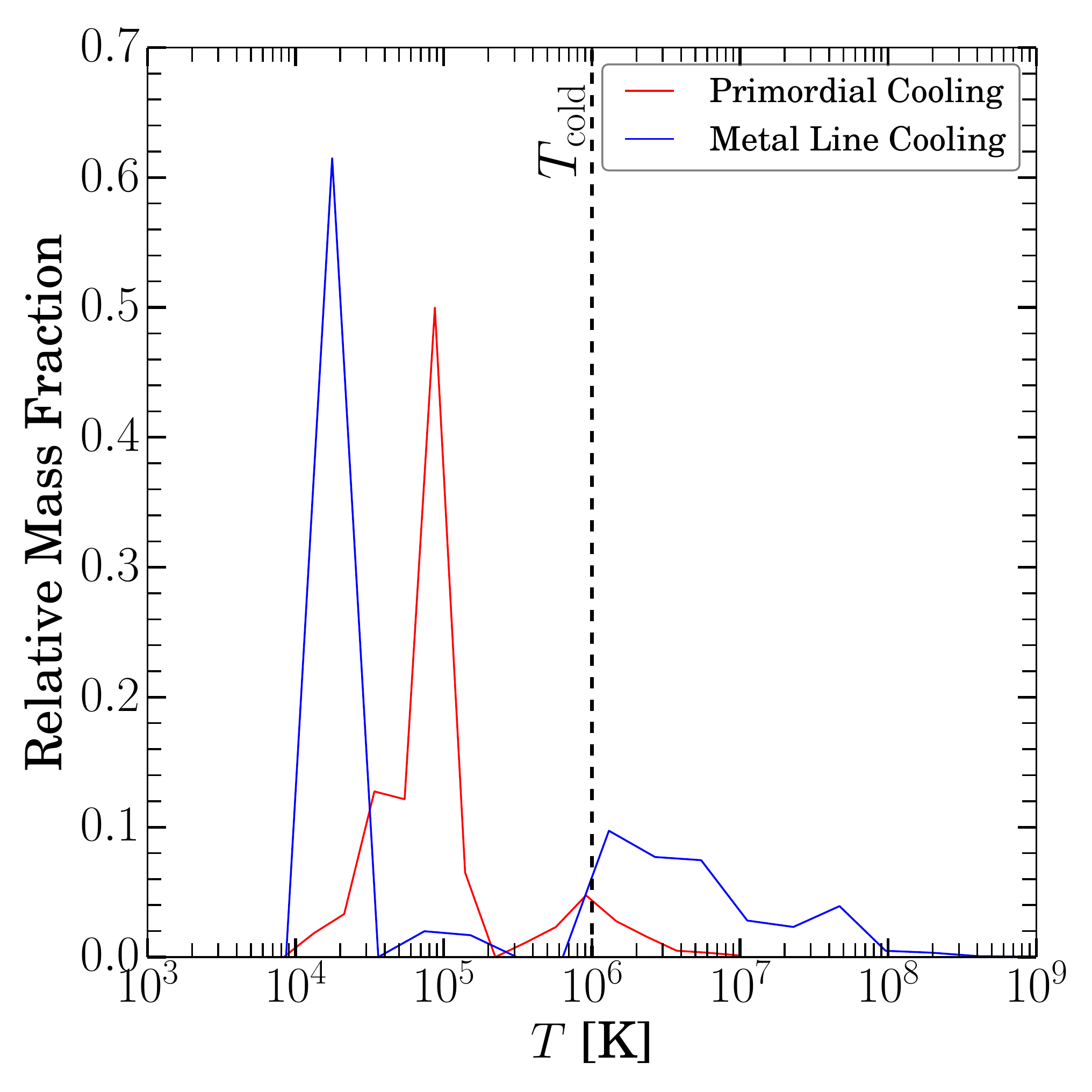}
\includegraphics[width=0.3\textwidth]{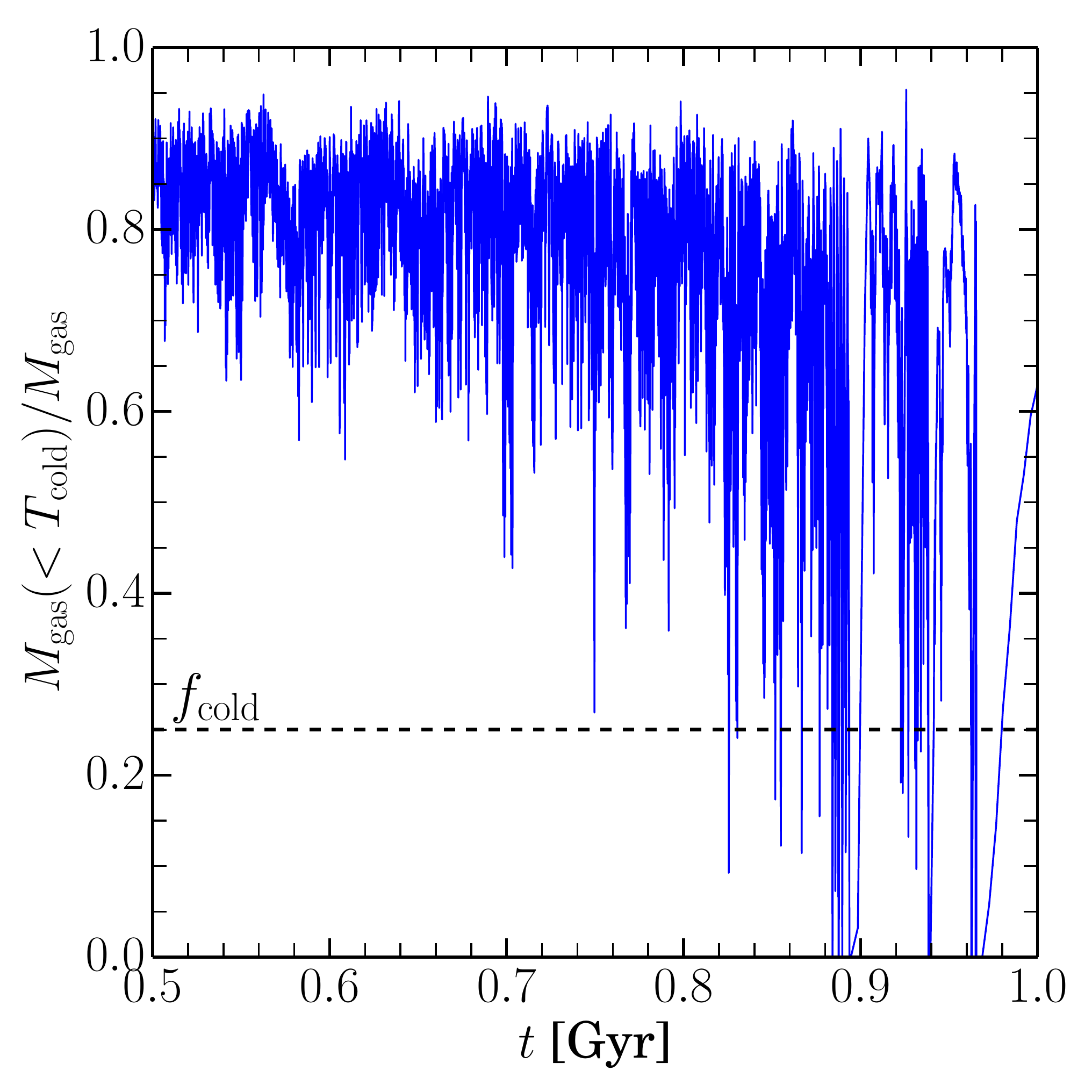}
\includegraphics[width=0.37\textwidth]{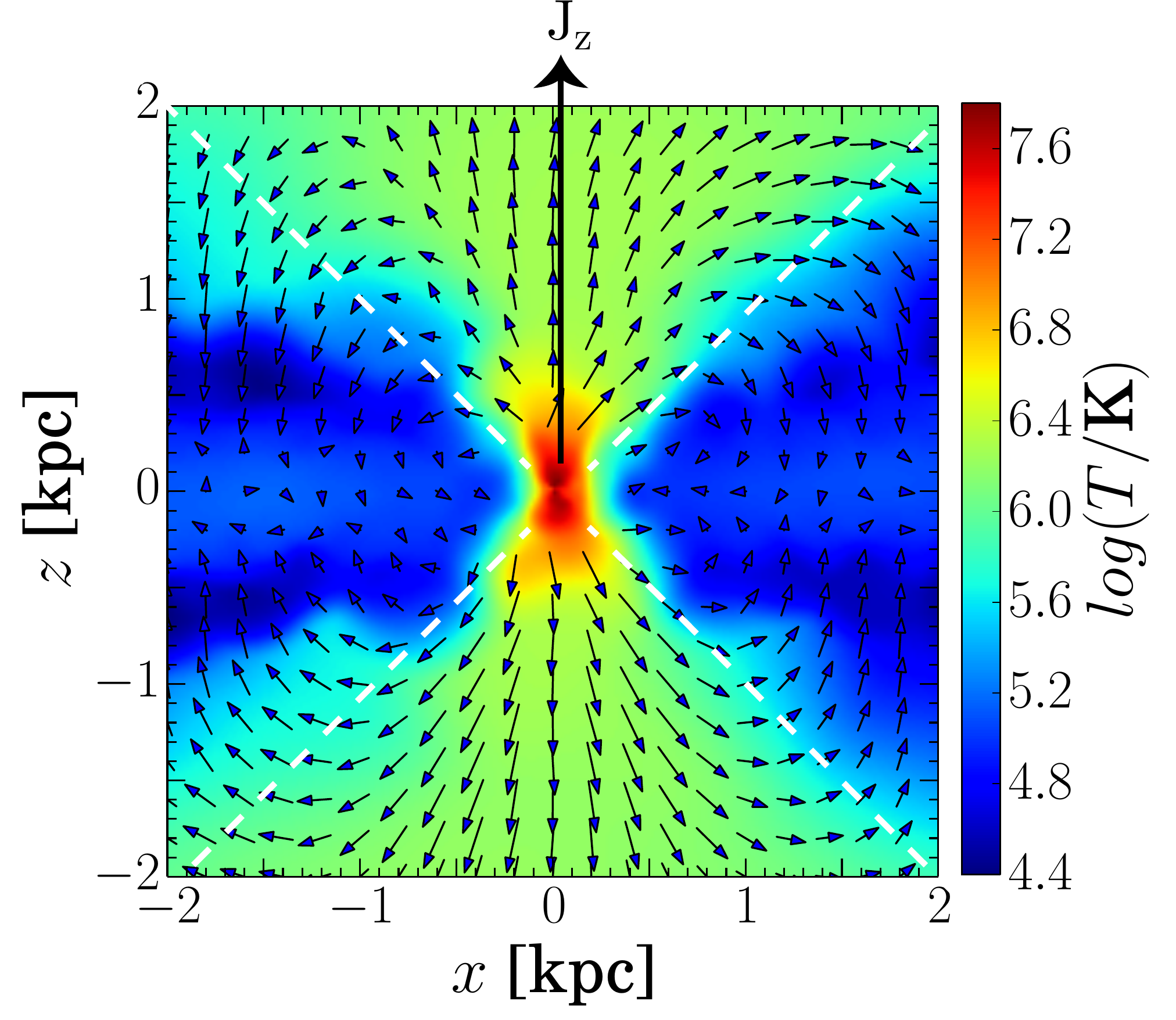}
\caption{In the left-hand panel, we show the temperature distribution of the
  gas within the black hole smoothing length, which is typically around $1\,{\rm kpc}$. Here we show the distribution
  for our normal isotropic thermal feedback model (red), and that for a
  similar simulation with metal line cooling (blue). The majority of the gas
  is within a clearly distinguished cold component at around $10^5$K (red) and
  $10^4$K (blue). The feedback heated gas is clustered around $10^6$K. In our
  bipolar model we measure the fraction of gas that is present in the cold
  component, which can be seen in the middle panel, and if the fraction is
  larger than $f_{\rm cold} = 0.25$ we estimate the properties of the accreting gas solely from
  that same cold component. In the right-hand panel we show gas temperature
  map using the bipolar feedback model. Overplotted arrows indicate the
  direction and relative velocity of the gas, and we also show the feedback
  cone (dashed white lines) and angular momentum axis of the accreting gas,
  $\vec J_{\rm z}$.}
\label{bi_mod} 
\end{figure*}

\subsubsection{Non isotropic accretion and feedback}

The traditional approach
of estimating the fluid parameters and injecting the feedback energy via a
simple, isotropic algorithm fails to make use of the increased resolving power
enabled by our super-Lagrangian refinement technique. To overcome this, in
this paper we explore the effects of a non-isotropic scheme on our black hole
model.

In certain physical configurations (e.g. efficient cooling of gas with
significant angular momentum) accreting material should consist of a
cold, dense disc component as opposed to the hot, spherically symmetric
gas. We attempt to model this by defining the cold component of the gas within the smoothing
length of the black hole as that with a temperature below
$T_\mathrm{cold}$. Whenever this cold component makes up a
fraction of at 
least $f_\mathrm{cold}$ of the mass of the gas within the
smoothing length, we 
enter a non isotropic mode of accretion. Here we estimate the fluid parameters
(i.e. $c_\infty$, $\rho_\infty$) only from the cold component, otherwise we
proceed as before in taking all gas within the smoothing length\footnote{Note
  that with this model we only explore how inflowing gas geometry affects our
  accretion model rather than attempting to self-consistently track the black hole
accretion disc, which occurs below our resolvable scale.}. For
consistency, when adopting this method we also limit the draining of mass from
the gas cells surrounding the black hole to those cells in the disc
component. There are clearly more sophisticated ways of accurately isolating
the cold component of the gas, but we wish to keep our method as simple as
possible and to make as few assumptions as possible, since this will allow us
to apply it to a cosmological context, where gas can have a very
complex morphology.

We also adapt our feedback algorithms to implement a non isotropic
scheme. Here, we are motivated by unresolved dynamics that lead to observed
bipolar outflows from AGN \citep{Rupke:11, Maiolino:12}. These may be driven by
magnetic fields close to the black hole that we do not attempt to model in
these simulations, or could be hydrodynamically driven on larger scales. Instead, we allow for their effects by implementing both
the thermal and momentum schemes detailed above, limiting them to a double
cone of opening angle $\theta_\mathrm{out}$ with axis parallel to the axis of
the angular momentum vector of the gas around the black hole which is an
  assumption of our model \citep[see
  also][]{Ostriker:10,Choi:14}. In addition, we also allow 
the feedback to entrain mass in the outflow. We specify an efficiency
parameter $\epsilon_\mathrm{out}$ which represents the
fraction of accreting 
gas that is swept up into the outflow. We inject the resulting mass into gas
cells within the double cone, weighting by the cubic spline kernel, in
an exact reverse of our draining procedure.

\subsubsection{Parameter choices}

We now describe how we set values for the free parameters, $T_{\rm cold}$,
$f_{\rm cold}$, $\epsilon_{\rm out}$ and $\theta_{\rm out}$ that define how
black holes accrete from 
the cold component of the gas. Left-hand panel of Figure~\ref{bi_mod} shows an
example of a typical distribution of the gas temperature for our isotropic
thermal feedback scheme, as well as for a similar simulation but with
additional metal line cooling. A clearly distinct cold component can be seen
around $10^5$K in the standard feedback model and at around $10^4$K in the case of metal line cooling. This is typical of the values that we see across our simulations (other than when the gas of the galaxy is exhausted at late times), and motivates using a temperature $T_\mathrm{cold}=10^6$K. Increasing this temperature  to values above $10^7$K recovers the results of our isotropic simulations.

Similarly, we set $f_\mathrm{cold}$, the fraction of the mass of the gas that
needs to be below the critical temperature in order for us to estimate the
accretion parameters from the cold component, to $0.25$. We have investigated
different choices of this parameter, but for values $< 0.4$ we find that the
results do not change noticeably. This can be explained by the middle panel of
Figure~\ref{bi_mod}, which shows the typical evolution of the fraction of cold
gas prior to the destruction of the central region of the disc (which happens
at late times due to 
build up of feedback energy). Here we see that until $t \sim 0.8 \, {\rm Gyr}$, the mass
of accreting gas present is well above our threshold, rendering the simulation
insensitive to small changes in this value. At later times, as the cumulative
feedback energy becomes significant, the cold fraction drops until it
eventually reaches the range when we switch to isotropic feedback. However,
when this happens, the rate of change of the cold fraction is very fast -
beyond a certain point, the cold disc component is quickly overcome. As such,
changes in $f_\mathrm{cold}$ result in the model switching to isotropic
feedback at nearly identical times.  

We set $\theta_\mathrm{out}$, the opening angle of the bipolar
outflow, to be $\pi/4$. In test simulations, we find that for a broad range of
opening angles, the accretion rate is consistent. This is perhaps unsurprising,
since in this case we estimate the fluid parameters from the cold component
and not from the hot outflow, and the majority of the accreting gas is centred
in the plane perpendicular to the angular momentum axis. We set the efficiency
of the entrainment of the gas into the outflow to
$\epsilon_\mathrm{out}=0.5$. This in principle may be much higher -
\citet{Choi:14} for example, find that mass and energy conservation (based
upon an outflowing wind velocity of $10000\, {\rm km \,s}^{-1}$ and the same
feedback efficiency used here) imply that $90\%$ of the in-falling gas will be
expelled in a wind. Our choice thus represents a more conservative value; a
full study of the impact of this parameter will follow in a future
paper. Right-hand panel of Figure~\ref{bi_mod} shows the result of this
bipolar model in practice. Here we plot a temperature map through the $x-z$
plane of the simulation, with the black hole centred at the origin. The
imprint of the bipolar outflow is visible in the centre, with the gas strongly
heated to almost $10^8$K. 

We also check that our measurement of the gas angular momentum (and as such the
direction along which we inject feedback) is robust. This is especially
important as any fluctuation in the axis of the feedback cone could lead to
the disc being destroyed prematurely. We confirm that in our simulations of
isolated galaxies, for which the net angular momentum axis should align with
the $z$ axis, whilst the cold disc is present, our estimated angular
momentum for the surrounding gas is consistently aligned with the $z$ axis. We further
verify that maintaining a fixed angular momentum vector along this axis does
not change the results.

\subsubsection{Duty cycle}

In addition to injecting feedback at each time step, we also investigate the
effects of allowing the feedback energy to build up over a time-scale
$t_\mathrm{cycle}$. Note that here, the time-scale defined is the time between
feedback events, not the length of the burst. After this time has elapsed, the
accreted energy is released in a single feedback event, and the cycle
repeats. This mimics the observed characteristics of AGN duty cycles,
but the time-scales themselves are poorly constrained. For the purposes of this
paper, we are interested in the effects on the sound speed of the gas, since
this will affect black hole accretion rate. A full parameter study is beyond
the scope of this paper - here we set $t_\mathrm{cycle}$ to $10^8\, {\rm yr}$ as it
was found that this was long enough to have a substantial impact of the gas
properties, but short enough to allow a significant number of accretion events
across the full time span of our simulations.

\section{Results: Bondi Inflows}

\label{Results:Bondi}

\subsection{Initial Conditions}

As a first test of our refinement technique we verify it's ability to improve
the resolution of the fluid flow in the region around the black hole. For this
purpose we carry out a series of simulations, where the black hole is surrounded
by a spherically symmetric gas distribution and where there is no net bulk
relative velocity between the gas and the black hole. By considering these
idealized initial conditions we can directly compare our measure of gas
accretion on to the central object with the theoretical Bondi rate, and examine
how this changes when our refinement scheme is used. 
  
We generate initial conditions based on the theoretical density and velocity
profiles of spherically symmetric accretion on to a point mass
\citep{Bondi:52}. The gas is characterized by its properties at infinity,
where it is at rest and has a uniform density $\rho_\infty$ and pressure
$p_\infty$. Conservation of mass gives the equation 
\begin{equation}\label{beq1}
\dot{M} = -4\uppi r^2\rho v,
\end{equation}
where $\dot{M}$ is the constant mass accretion rate. Bernoulli's equation then gives
\begin{equation}\label{beq2}
\frac{1}{2}v^2 + \left(\frac{\gamma}{\gamma-1}\right)\frac{p_\infty}{\rho_\infty}\left[\left(\frac{\rho}{\rho_\infty}\right)^{\gamma-1} - 1\right] = \frac{GM_{\mathrm{BH}}}{r},
\end{equation}
where we have assumed a Newtonian gravitational potential for the central supermassive black hole, as well as a polytropic equation of state such that $p/p_\infty = (\rho/\rho_\infty)^\gamma$, for $1 \le \gamma \le 5/3$.

The accretion rate is then given by
\begin{equation}
\dot{M} = \frac{4\uppi \lambda\, G^2 M_\mathrm{BH}^2 \rho_\infty}{c_\infty^3} ,
\end{equation}
where $\lambda$ is a non-dimensional parameter equal to \citep{Bondi:52}
\begin{equation}
\lambda = \left(\frac{1}{2}\right)^\frac{\gamma  + 1}{2(\gamma-1)}\left(\frac{5-3\gamma}{4}\right)^\frac{3\gamma-5}{2(\gamma-1)}.
\end{equation}

We solve Equations \ref{beq1} and \ref{beq2} for the two unknowns, $\rho(r)$
and $v(r)$, which we hereafter refer to $\rho_\mathrm{B}(r)$ and
$v_\mathrm{B}(r)$, respectively. We then use these as the basis for generating
our initial conditions.

We use a near identical setup to that of \citet{Barai:11}: we generate a spherical
distribution of gas surrounding a central supermassive black hole with mass
$M_\mathrm{BH} = 10^8 \mathrm{M}_\odot$. We distribute the gas by randomly sampling
from the cumulative mass distribution given by $\rho_\mathrm{B}(r)$, and then
by setting the appropriate radial velocity $v_\mathrm{B}(r)$. The gas is
distributed between an inner radius $r_\mathrm{in}$ and an outer radius
$r_\mathrm{out}$ and we set $T_\infty = 10^7\, {\rm K}$ and $\rho_\infty = 10^{-19}
\, {\rm g\, cm^{-3}}$. We set $\gamma = 1.01$ which, coupled with a value of $r_\mathrm{in}
= 0.1\, {\rm pc}$ and $r_\mathrm{out} > 10\, {\rm pc}$ means that both $r_{\rm
  B} = 3.0\, {\rm pc}$ and
$r_\mathrm{sonic}$ = $1.5\, {\rm pc}$ (the point at which the in-falling gas becomes supersonic) lie well within our simulation domain.

\begin{figure*}
\centering
\includegraphics[width=0.45\textwidth]{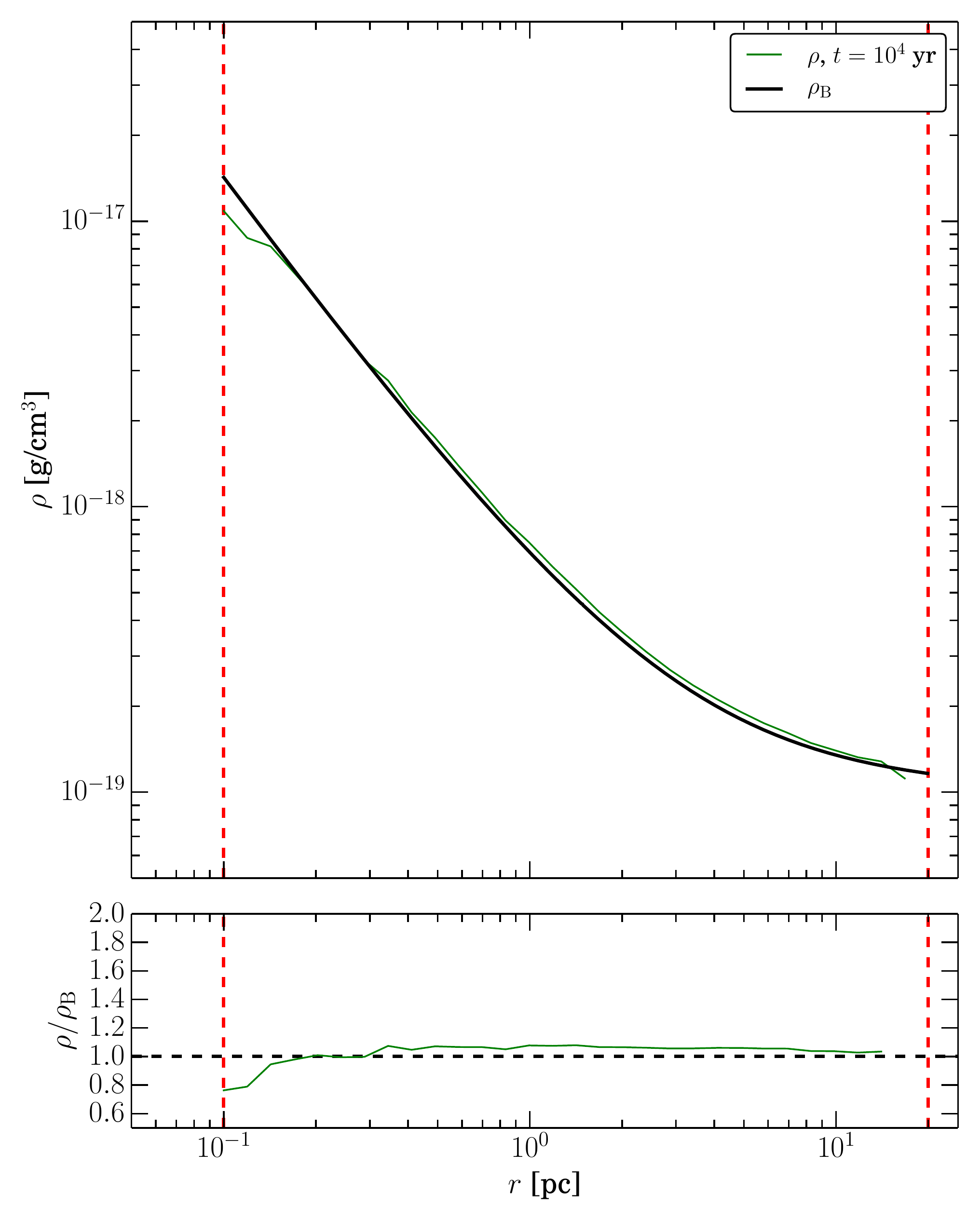}
\includegraphics[width=0.45\textwidth]{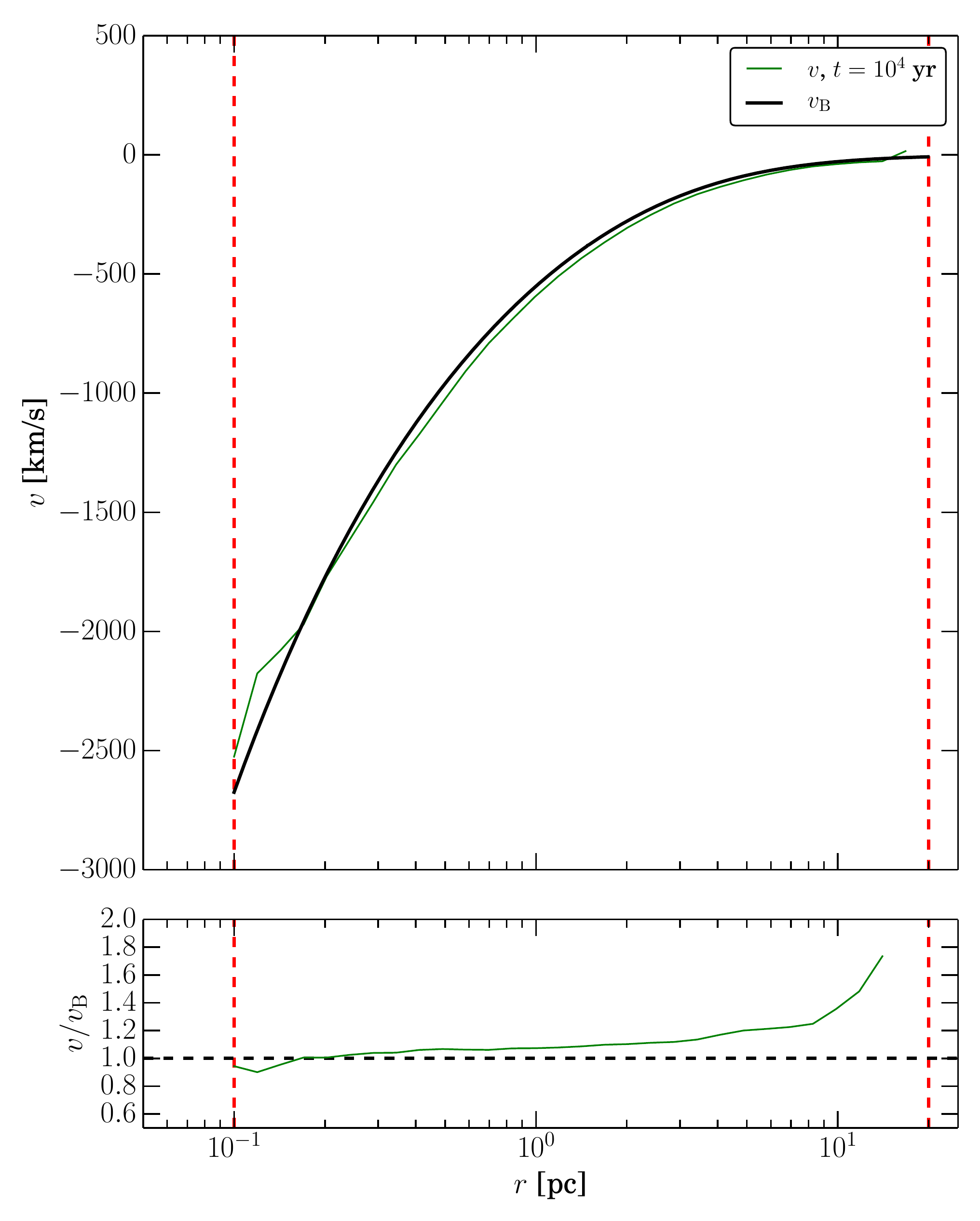}
\caption{Radial profiles of gas density (left-hand panel) and gas velocity
  (right-hand panel). The black curves denote the analytical Bondi solution while the green
curves are our simulation results at $t = 10^4\, {\rm yrs}$ for our simulation without refinement, with $64^3$ initial resolution elements. Red, vertical dotted lines indicate
$r_\mathrm{in} = 0.1\, {\rm pc}$ and $r_\mathrm{out} = 20\, {\rm pc}$. In the lower panel, we show the ratio of the analytic profile with the measured one. Small departures from the
analytical solution at radii close to $r_\mathrm{in}$ occur due to the inner boundary condition
imposed there, but note that their effect is negligible given that the Bondi
radius is located at $r_{\rm B} = 3.0\, {\rm pc}$.}
\label{bondi_profiles}
\end{figure*}

\begin{table*}
\bc
\begin{tabular}{ccccccccc}
\hline\hline
\textbf{Name} & $N_\mathrm{gas}$ & $r_\mathrm{out}$ (pc) & Refinement &
$R_\mathrm{ref}$ (pc) & $R_\mathrm{cell}^\mathrm{min}$ (pc) &
$R_\mathrm{cell}^\mathrm{max}$ (pc) & CPU time (h)\\ \hline
NoRef $16^3$ & $16^3$ & 20 & No & - & - & - & 4.4\\
NoRef $32^3$ & $32^3$ & 20 & No &- & - & - & 12.3\\
NoRef $64^3$ 10 pc & $64^3$ & 10 & No & - & - & - & 102.4\\
NoRef $64^3$ 15 pc & $64^3$ & 15 & No &- & - & - & 91.3\\
NoRef $64^3$ & $64^3$ & 20 & No & - & - & - & 75.2\\
Ref $16^3$ & $16^3$ & 20 & Yes & 1.0 & $0.01$ & $1$ & 5.0\\
Ref $32^3$ & $32^3$ & 20 & Yes & 1.0 & $0.01$ & $1$ & 14.4\\
RefAggr $32^3$ & $32^3$ & 20 & Yes & 1.0 & $0.001$ & $0.5$ & 54.6\\
\hline\hline
\end{tabular}
\caption{Simulation details of Bondi inflow models. First column lists the
  name of simulations performed, second column indicates the initial number of gas
  cells used, while third column gives the outer radius of the simulation
  domain beyond which gas conditions at infinity are imposed. In the fourth
  column we indicate if the super-Lagrangian refinement scheme is used or not,
  and for the subset of simulation with refinement on, we list the refinement
  parameters, namely $R_{\rm ref}$, $R_\mathrm{cell}^\mathrm{min}$ and
  $R_\mathrm{cell}^\mathrm{max}$ in columns five, six and seven,
  respectively. Finally, in column eight we show the total CPU hours used by $t \,
  =\, 6 \times 10^4\, {\rm yrs}$.}
\label{tb:bondisims} 
\ec
\end{table*}

\subsection{Numerical setup}

\begin{figure*}
\centering
\includegraphics[width=0.33\textwidth]{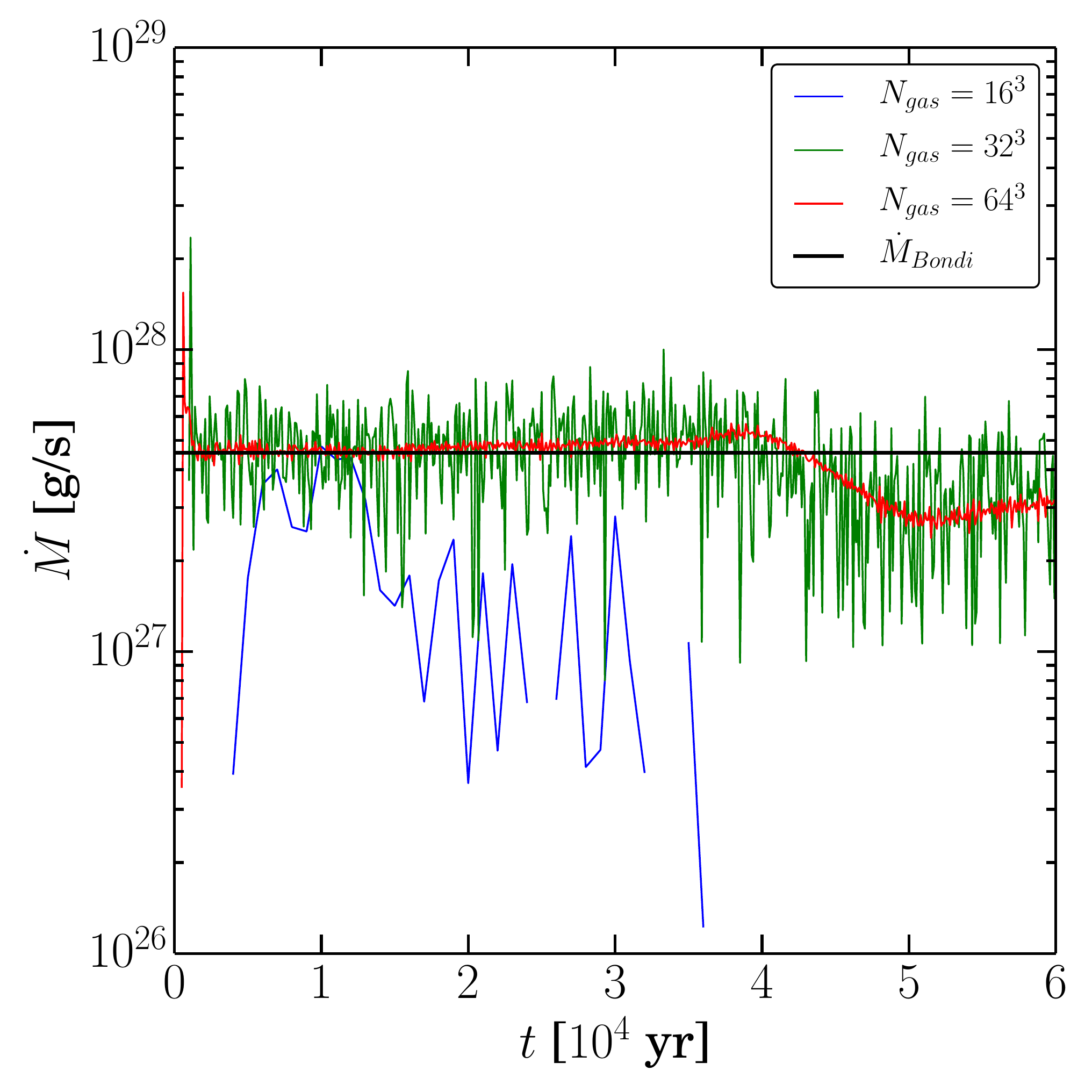}
\includegraphics[width=0.33\textwidth]{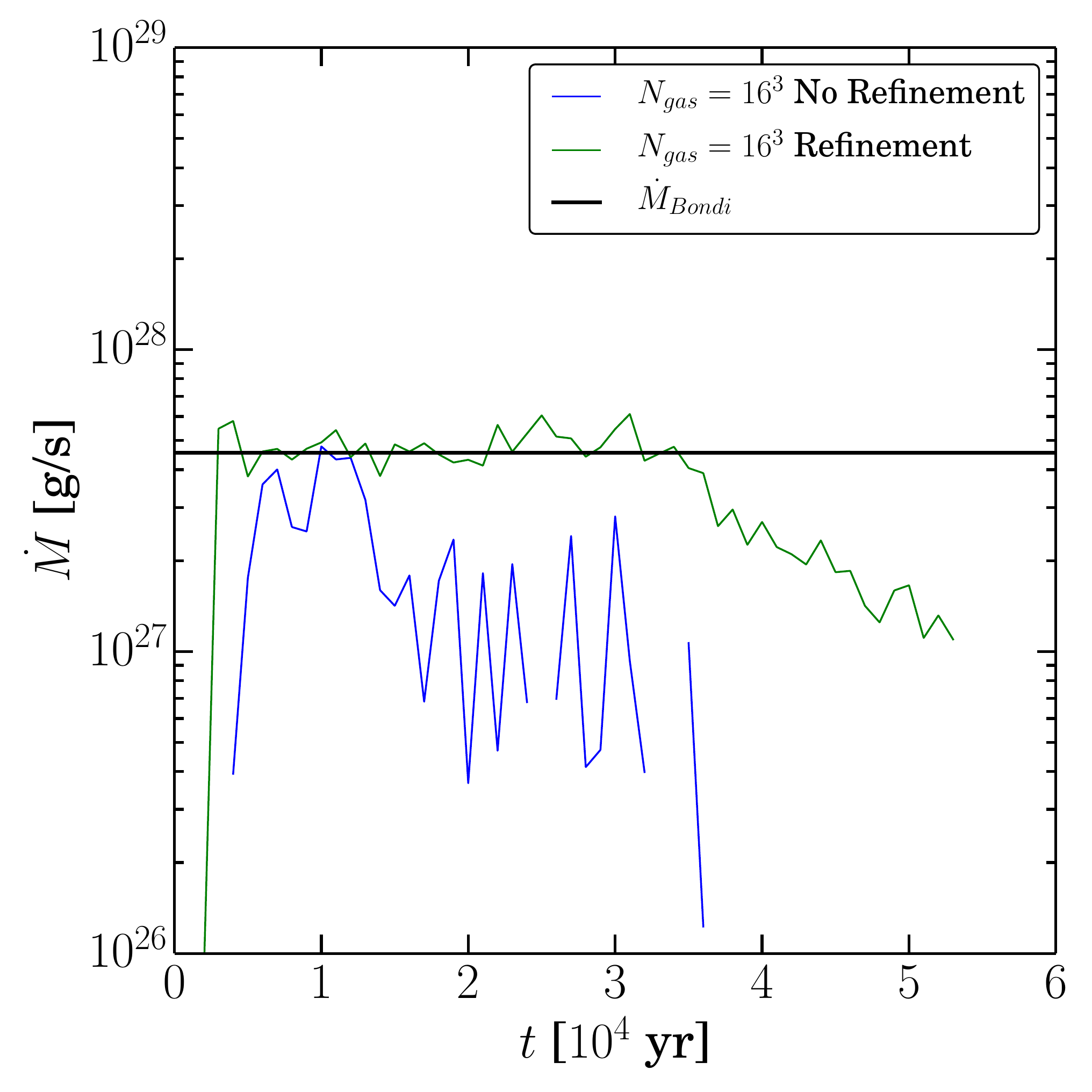}
\includegraphics[width=0.33\textwidth]{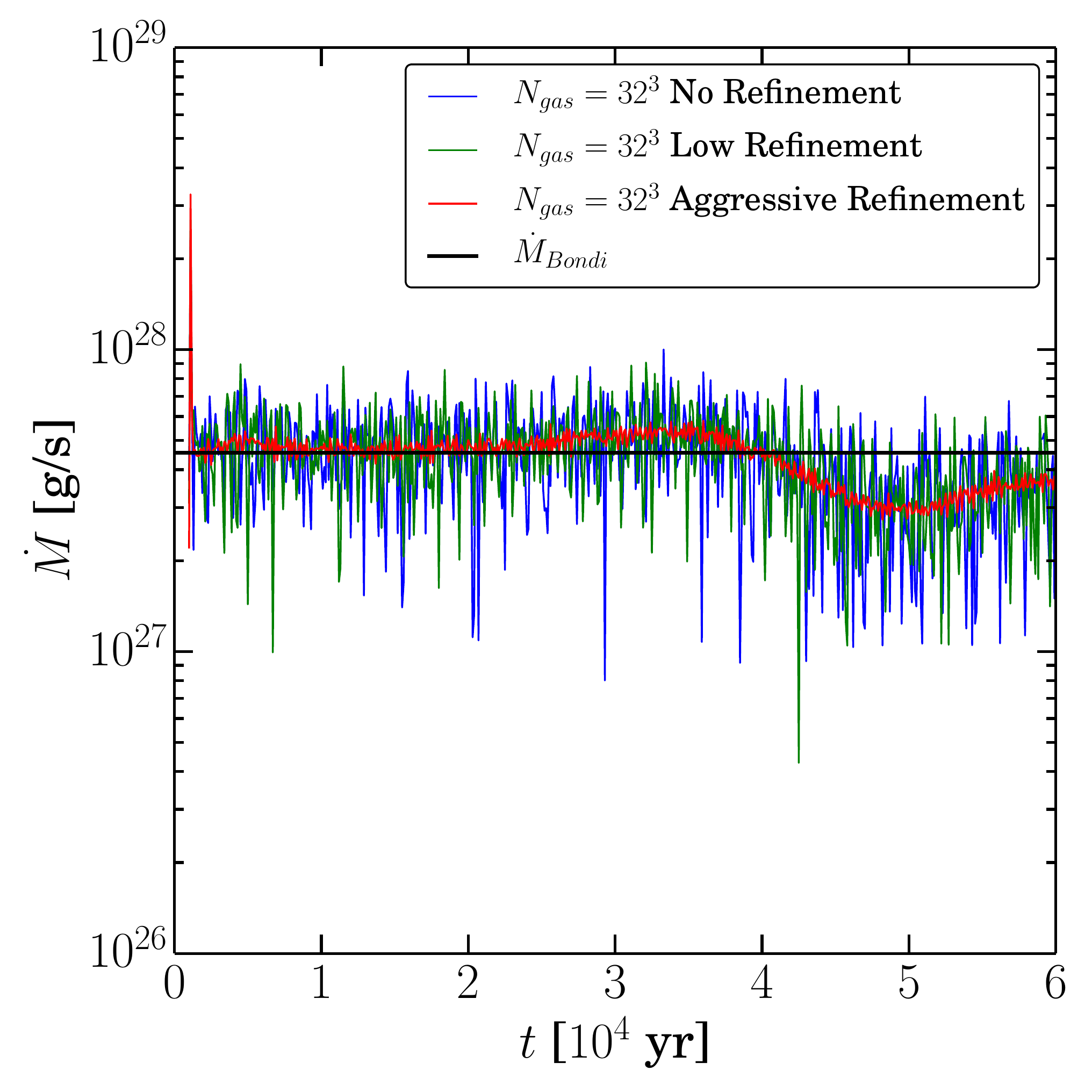}
\caption{Mass inflow rate across $R = 0.1\, {\rm pc}$ averaged over a fixed size
  timestep $\mathrm{d}t$ for simulations with $N_{\rm gas} =  16^3, 32^3$ and $64^3$
  gas cells ($\mathrm{d}t = 1000\, {\rm yr}$ for $N_{\rm gas} =  16^3$, $\mathrm{d}t = 100\,
  {\rm yrs}$
  otherwise). The analytic solution is indicated by the horizontal solid black line. Left-hand
  panel: results without our super-Lagrangian refinement for 
  three different resolutions as shown on the legend. Central panel:
  comparison of the results without and with black hole refinement at a
  fixed resolution of $N_{\rm gas} =  16^3$. Right-hand panel: no, moderate
   and aggressive refinement at a fixed resolution of $N_{\rm gas} =  32^3$. Note
that the discrepancy between the analytical prediction and simulation results
for $t \gtrsim 4 \times 10^4\, {\rm yr}$ is entirely due to the location of the outer
boundary condition (i.e. the size of the simulated region; see text for more details).}  
\label{ref_bondi}
\end{figure*}

We make some minor adjustments to our code in order to enforce the boundary
conditions of the theoretical setup. We ensure that gas cells situated at a
distance of more than $r_\mathrm{out}$ from the centre of the simulation have
the parameters of the gas at infinity - i.e. $T_\infty$, $\rho_\infty$,
$p_\infty$ - and that they are stationary. In principle this means there is a
small discontinuity in the density distribution at $t=0$ that decreases with
larger values of $r_\mathrm{out}$. We examine the effects of this in the
section below. In these simulations, we replace our usual estimate of the accretion based on the large scale fluid properties with a routine that swallows individual gas cells. In the centre of the simulation domain, we use a simple sink
particle routine to remove gas cells from the region closest to the black
hole. Here, we remove all
cells that have both mesh generating points and 
estimated semi-major axes (of their orbit around the black hole) within $r_\mathrm{in}$. We do, however, keep the mass
of the central black hole fixed to maintain a constant theoretical accretion
rate with time. This does not substantially affect the results, as the total
accreted mass throughout our simulations represents at most about 5\% of
$M_\mathrm{BH}$. We sum the mass of all cells removed at the inner radius and
take this as our estimate of the accretion rate. For completeness, we also
calculate the total mass of cells within the Bondi radius. We check that
this remains constant throughout the simulation, indicating that (as expected)
the flux across the Bondi radius is the same as that across the inner radius
and that for sufficiently small radii, the choice of $r_\mathrm{in}$ does not
affect our results.

We run a suite of simulations (see Table~\ref{tb:bondisims}), varying both the
number of gas cells $N_\mathrm{gas}$ and the outer truncation radius
$r_\mathrm{out}$ for simulations with and without refinement. For simulations
in which the refinement scheme is active, we also examine the effects of more
aggressive refinement (by changing the refinement parameters
$R^\mathrm{cell}_\mathrm{min}$ and $R^\mathrm{cell}_\mathrm{max}$).

In Figure~\ref{bondi_profiles} we plot the radial profiles of gas density (left-hand panel) and gas velocity
  (right-hand panel). Black curves denote analytical Bondi solution while green
curves are our simulation results at $t = 10^4\, {\rm yrs}$ (NoRef $64^3$). The simulation results agree very well with
the analytical profiles over the whole simulated timespan and over the
simulated spatial domain. We note that there are small departures from the
analytical solutions close to the inner, $r_\mathrm{in}$, and outer,
$r_\mathrm{out}$, boundary conditions. Their effect is however very small (as
we have explicitly checked by varying the values of $r_\mathrm{in}$ and
$r_\mathrm{out}$) given that the Bondi radius is located at $r_{\rm B} = 3.0\, {\rm pc}$.   

\subsection{Impact of resolution and refinement method}

In the left-hand panel of Figure~\ref{ref_bondi} we show the results of our simulations without
refinement for three different resolutions: $16^3$, $32^3$ and $64^3$ initial
gas cells. We also plot the theoretical accretion rate
$\dot{M}_\mathrm{B}$ denoted with the horizontal black line. For the highest
resolution case we find excellent 
agreement with the theoretical rate, with the exception for $t \gtrsim
4 \times 10^4\, {\rm yrs}$. This coincides with the sound travel time from our
truncation radius to the centre of the simulation domain, and this is the effect of
the small discontinuity in our initial conditions. We confirm this by checking
the accretion rate for simulations with successively smaller values of
$r_\mathrm{out}$ and find a linear relation between truncation radius and the
feature seen here. The simulations are otherwise identical, and after the wave
has propagated the simulations return to the theoretical rate.

For the medium resolution case, we see that there is still good agreement with
the theoretical picture, but with heightened stochasticity due to the Poisson
noise. By the time we get to our lowest resolution simulation, however, the
resolution is too coarse to correctly model the Bondi inflow and we do not
match $\dot{M}_\mathrm{B}$. 

The central panel of Figure~\ref{ref_bondi} shows an identical setup with $16^3$ initial gas cells
with and without our refinement technique. Here we see that the simulation
with super-Lagrangian refinement does a much
better job - even with the low resolution initial conditions we are able to
reproduce the theoretical Bondi accretion rate and the total CPU time used
lies in between the costs for the $16^3$ and $32^3$ runs without refinement. 

Furthermore, right-hand panel of Figure~\ref{ref_bondi} shows how changing the
aggression of the refinement parameters affects the simulated accretion
rate. Here we show the medium resolution case with initial $32^3$ gas
cells. With moderate refinement, the stochasticity of the cell flux over
the inner radius is reduced, but only slightly. For more aggressive parameters
this is dramatically reduced and we are able to reproduce an accretion rate
that is nearly identical to the run with eight times the number of  cells
at somewhat lower CPU cost.

\section{Results: Isolated disc galaxy simulations} \label{Results:Galaxy}
\label{Results:Ref}

\begin{table*}
\bc
\begin{tabular}{ccc}
\hline\hline
\textbf{Parameter} & \textbf{Value} & \textbf{Description} \\ \hline 
c & 9.0 & Halo concentration \\ 
$v_{200}$ & 160 $\mathrm{km\, s^{-1}}$ & Virial velocity\\
$M_{200}$ & $9.52\times 10^{11} M_{\mathrm \odot}$ & Virial mass\\ 
$m_\mathrm{d}$ & 0.041 & Disc mass fraction\\ 
$m_\mathrm{b}$ & 0 & Bulge mass fraction\\ 
$h$ & 2.74 $\mathrm{kpc}$ & Disc scalelength\\
$z_0$ & $0.2 \,h$ & Disc scaleheight\\ 
$\lambda$ & 0.033 & Spin parameter\\ 
$m_\mathrm{BH}$ & $10^5 - 10^{6} M_{\mathrm \odot}$ & Black hole seed mass \\ 
$f_\mathrm{gas}$ & $0.1$ & Gas mass fraction of the disc \\
$J_\mathrm{d}$ & $0.041$ & $m_\mathrm{d} \times$ total angular momentum of the halo \\\hline\hline
\end{tabular}
\caption{Parameter choices for the isolated galaxy models.}\label{tb:params}
\ec
\end{table*}

\subsection{Initial conditions} \label{iso}
\begin{figure}
 \includegraphics[width=0.99\columnwidth]{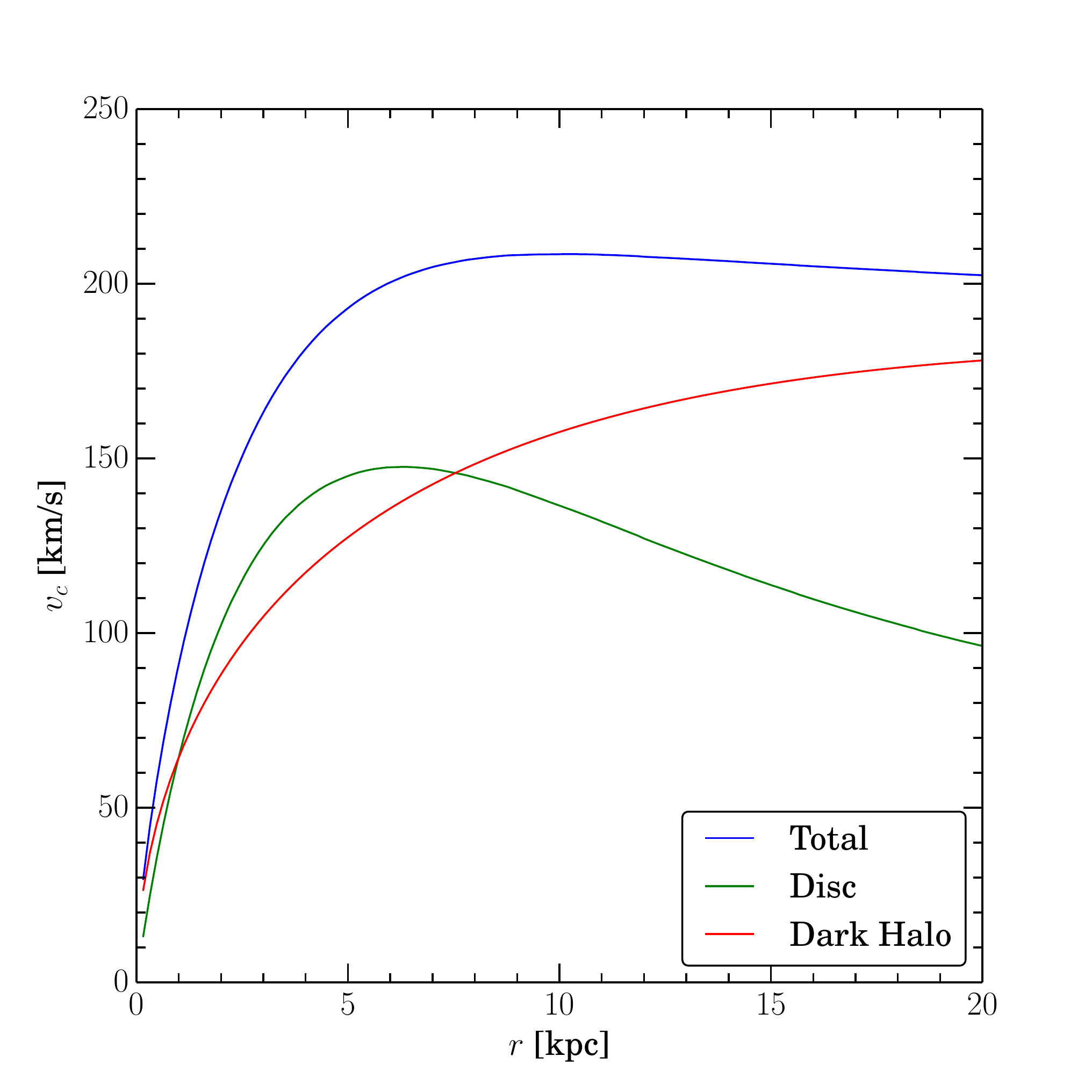}
 \caption{Rotation curves for our model galaxies, with parameters as detailed
   in Table \ref{tb:params}. The red line shows the circular velocity as a
   function of radius for the dark matter halo, whilst the green line shows the
   same for the disc component. The blue line shows the total rotation
   curve.}
 \label{velo_curve}
\end{figure}

We investigate the effects of our new refinement scheme using simple models of
isolated galaxies, as described by \citet{Springel:05}. Each galaxy consists
of a dark matter halo with a \citet{Hernquist:90} density profile 
\begin{equation}
\rho_\mathrm{dm} = \frac{M_\mathrm{dm}a}{2\uppi r(r+a)^3}\;,
\end{equation}
where $M_\mathrm{dm}$ is the total dark matter mass and $a$ is a scaling
parameter. This profile has the advantage that for inner radii it is similar
to the Navarro, Frenk and White \citep[NFW]{NFW} model but has a finite
mass. The two can be related by choosing  
\begin{equation}
a = r_\mathrm{s}\sqrt{2(\ln(1+c)-c/(1+c))}\;,
\end{equation}
where $c=r_{200}/r_\mathrm{s}$ is the concentration index of the NFW halo, and $r_\mathrm{s}$ is
the scale radius. Both gaseous and stellar discs are modelled using an
exponential surface density profile:
\begin{align}
\Sigma_\mathrm{gas}(r) &= \frac{M_\mathrm{gas}}{2\uppi h^2}\,\mathrm{e}^{-r/h} \;,\\
\Sigma_\star(r) &= \frac{M_\star}{2\uppi h^2}\,\mathrm{e}^{-r/h}\;,
\end{align}
where $h$ is the scalelength of the disc, $M_\mathrm{gas}$ is the total
initial gas mass and $M_\star$ is the total initial stellar mass.  By assuming
that the disc is centrifugally supported and that it is negligibly thin
compared to its scalelength, $h$, its angular momentum can be expressed by 
\begin{equation}
J_\text{d}=M_\mathrm{d}\int^\infty_0 v_\mathrm{c}(R)\left(\frac{R}{h}\right)^2\mathrm{e}^{-\frac{R}{h}}\mathrm{d}R\;,
\end{equation}
where $M_\mathrm{d}$ is the total mass in the disc, $R$ is the cylindrical radius and $v_\mathrm{c}$ is the circular velocity.

The vertical mass distribution of the stellar disc is given the profile of an
isothermal sheet, resulting in a three dimensional stellar density
distribution of 
\begin{equation}
\rho_\star(R,z)=\frac{M_\star}{4\uppi z_0h^2}\mathrm{sech}^2\left(\frac{z}{z_0}\right)\mathrm{e}^{-\frac{R}{h}}\;,
\end{equation}
where $z_0$ determines an effective temperature of the
disc. The velocity dispersion is then set to self-consistently maintain this
scaleheight in the full 3D potential of the model. 

The gaseous disc is governed by hydrostatic equilibrium, whereby the azimuthal
velocity is set to ensure balance between the inward gravitational and the
outward combination of centrifugal and pressure support 
\begin{equation}
v_{\rm \phi,gas}^2 = R\left(\frac{\partial \Phi}{\partial R} +
\frac{1}{\rho_{\rm gas}}\frac{\partial P}{\partial R}\right),
\end{equation}
with the remaining velocity components $v_R = v_z = 0$. An initial
distribution is found, which is used as the basis for an iterative method
which produces a self-consistent gas distribution and
potential. Table~\ref{tb:params} shows our parameter choices, which are
chosen to represent a simplified typical Milky Way-like disc galaxy, with the
rotation curves for different components shown in
Figure~\ref{velo_curve}. Table~\ref{tb:sims} lists the full suite of isolated disc
galaxy simulations that we have performed.

\begin{table*}
\bc
\begin{tabular}{ccccccccc}
\hline\hline
\textbf{Name} & \textbf{N} & \textbf{Refinement} & \textbf{Accretion} &
\textbf{Feedback} & ${m_\mathrm{target}}$ $(\mathrm{M}_\odot)$& $\epsilon_\mathrm{DM}$
$({\rm kpc})$ & $\epsilon_\mathrm{gas}$ $({\rm kpc})$ & \textbf{Cooling}\\ \hline
NoRef $10^5$ & $10^5$ & None & Default Bondi & Thermal & $2.0 \times 10^{5}$ &
2.0 & 0.15 & Primordial\\
NoRef $10^6$ & $10^6$ & None & Default Bondi & Thermal & $2.0 \times 10^{4}$ &
0.93 & 0.07 & Primordial\\
NoRef $10^7$ & $10^7$ & None & Default Bondi & Thermal & $2.0 \times 10^{3}$ &
0.43 & 0.032 & Primordial\\   
RefNorm $10^5$ & $10^5$ & Yes & Default Bondi & Thermal & $2.0 \times 10^{5}$
& 2.0 & 0.15 & Primordial\\
RefNorm $10^6$ & $10^6$ & Yes & Default Bondi & Thermal & $2.0 \times 10^{4}$
& 0.93 & 0.07 & Primordial\\
RefNorm $10^7$ & $10^7$ & Yes & Default Bondi & Thermal & $2.0 \times 10^{3}$
& 0.43 & 0.032 & Primordial\\ 
RefNorm $10^5$ & $10^5$ & Yes & Default Bondi & Momentum & $2.0 \times 10^{5}$
& 2.0 & 0.15 & Primordial \\  
RefNorm $10^5$ & $10^5$ & Yes & Default Bondi & Duty Cycle & $2.0 \times 10^{5}$ & 2.0 & 0.15 & Primordial \\ 
RefNorm $10^5$ & $10^5$ & Yes & Default Bondi & Bipolar & $2.0 \times 10^{5}$
& 2.0 & 0.15 & Primordial \\ 
RefNorm $10^5$ & $10^5$ & Yes & Default Bondi & Bipolar & $2.0 \times 10^{5}$ & 2.0 & 0.15 & + Metals \\
Fixed RefNorm $10^5$ & $10^5$ & Yes &
$\dot{M}_\mathrm{BH}=0.1\dot{M}_\mathrm{Edd}$ & Thermal & $2.0 \times 10^{5}$
& 2.0 & 0.15 & Primordial\\ 
Fixed RefNorm $10^5$ & $10^5$ & Yes &
$\dot{M}_\mathrm{BH}=0.1\dot{M}_\mathrm{Edd}$ & Momentum & $2.0 \times 10^{5}$
& 2.0 & 0.15 & Primordial\\  
Fixed RefNorm $10^5$ & $10^5$ & Yes &
$\dot{M}_\mathrm{BH}=0.1\dot{M}_\mathrm{Edd}$ & Duty Cycle & $2.0 \times
10^{5}$ & 2.0 & 0.15 & Primordial\\ 
Fixed RefNorm $10^5$ & $10^5$ & Yes &
$\dot{M}_\mathrm{BH}=0.1\dot{M}_\mathrm{Edd}$ & Bipolar & $2.0 \times 10^{5}$
& 2.0 & 0.15 & Primordial\\ 
Fixed RefNorm $10^5$ & $10^5$ & Yes & $\dot{M}_\mathrm{BH}=0.1\dot{M}_\mathrm{Edd}$ & Bipolar & $2.0 \times 10^{5}$ & 2.0 & 0.15 & + Metals\\ 
\hline\hline
\end{tabular}
\caption{Simulation details of isolated disc galaxy
  models. We list the name of the simulation that we use throughout this
  paper, as well as the initial number of resolution elements. Note that for
  all simulations, and especially those with refinement, the number of gas
  cells will increase from this initial value. In column 4 we indicate the
  accretion rate prescription we use, as well as the feedback mechanism in
  column 5. In columns 6-8 we list the target gas mass of the
  simulation (that is set to be the initial gas cell mass) as well as the dark
  matter and gas smoothing lengths. Note that in the course of simulations
  the gas gravitational softening is adaptive and is set to the maximum of
  $\epsilon_{\rm gas}$ and $2.5$ times the cell size. The last column indicates
  simulations where we include metal line cooling additionally to the
  primordial cooling.}\label{tb:sims} 
\ec
\end{table*}

\subsection{Validation of Implementation}

\begin{figure*}
\centering
\includegraphics[width=0.33\textwidth]{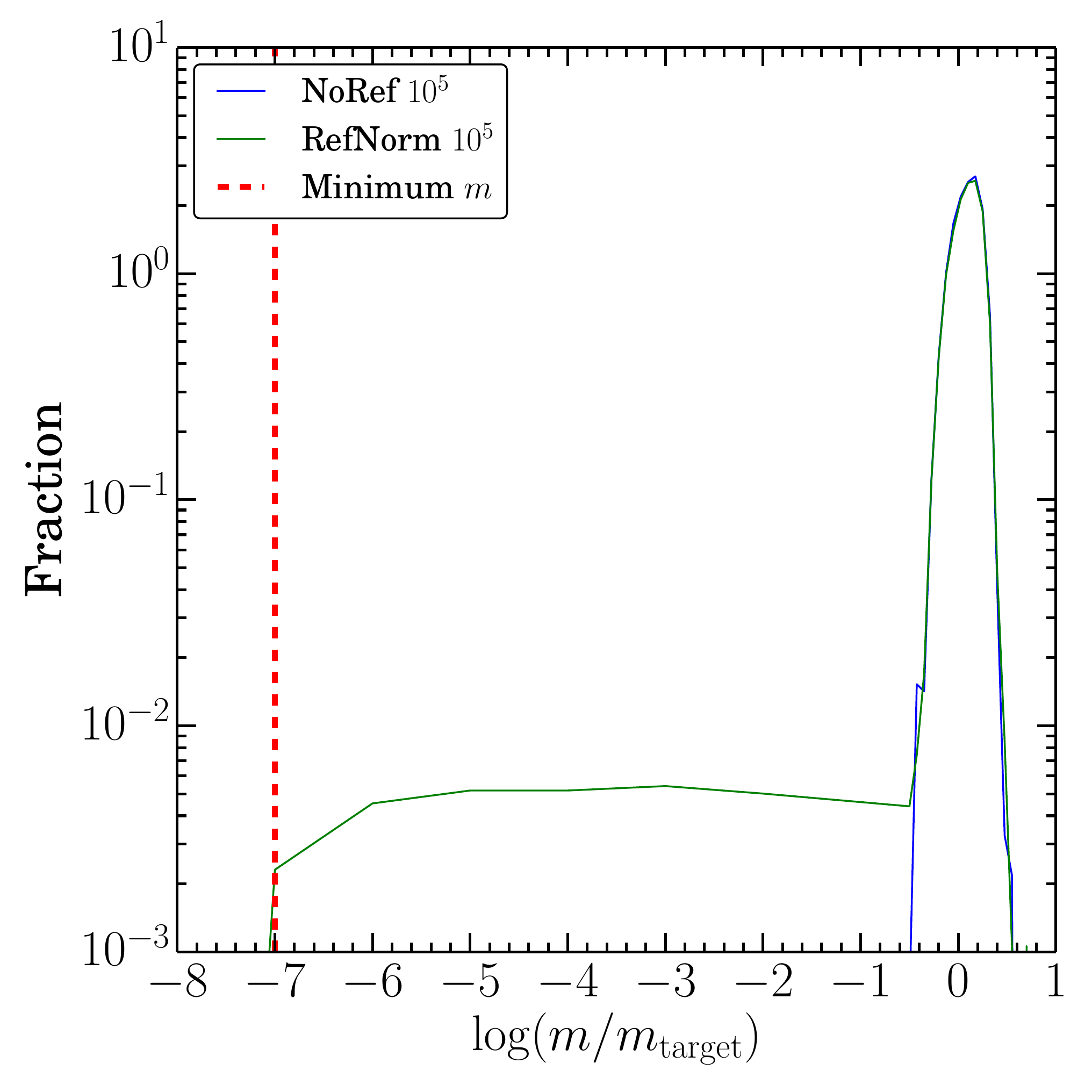}
\includegraphics[width=0.33\textwidth]{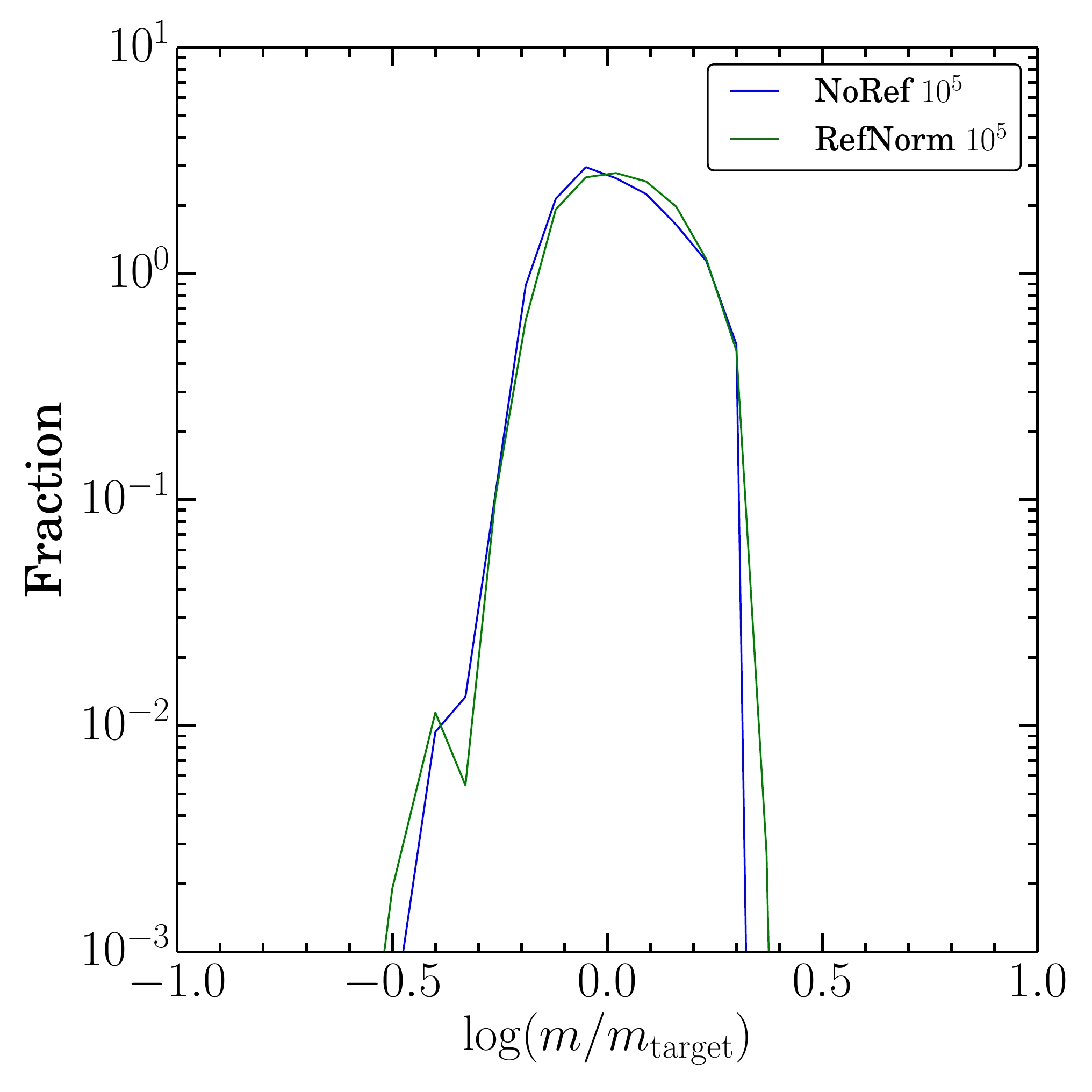}
\includegraphics[width=0.33\textwidth]{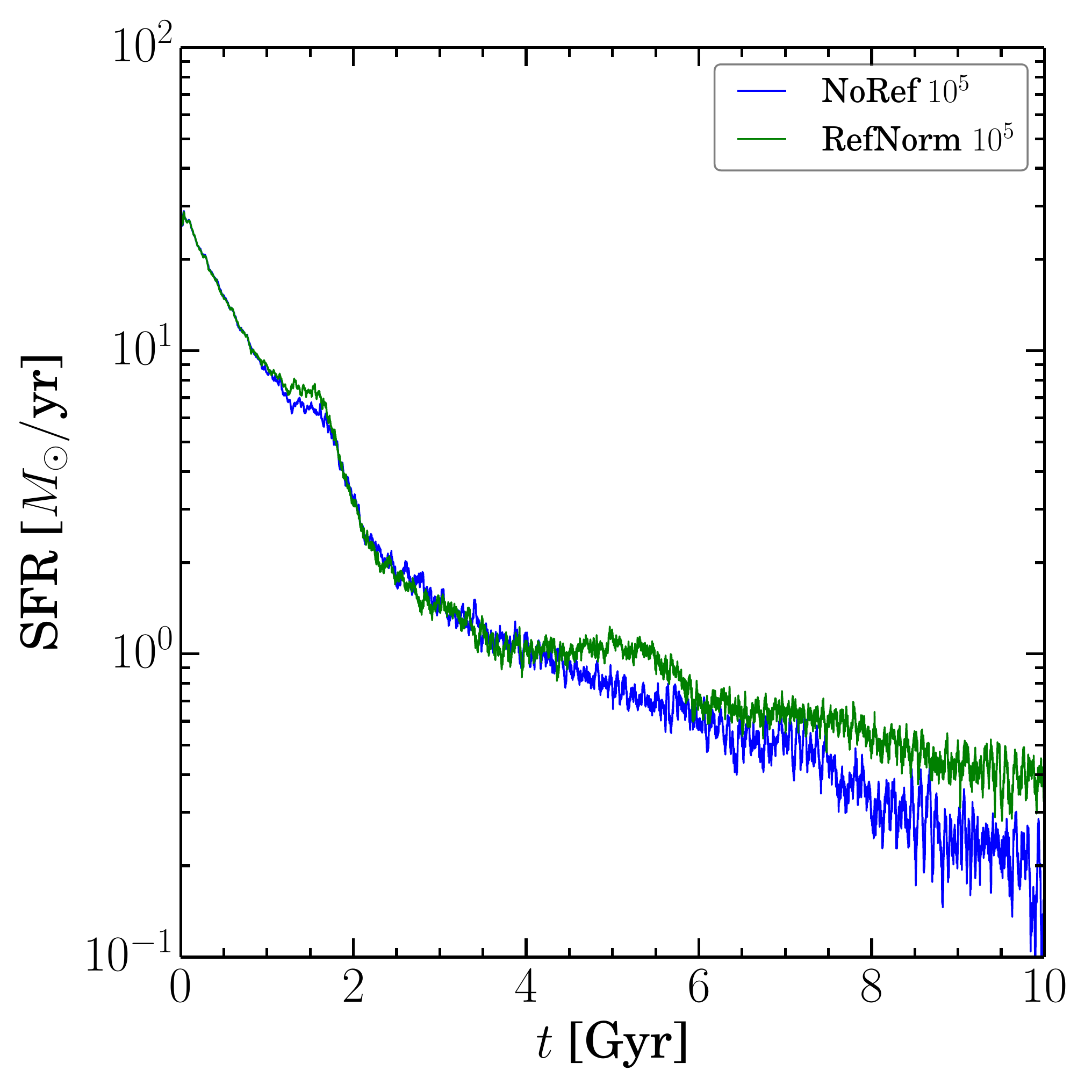}
\caption{Effects of the refinement technique. In the left-hand panel we plot the mass distribution of the gas
  cells, comparing that for our simulations without and with refinement. The
  distributions are in very good agreement at the high mass end, which
  demonstrates that we are not de-refining cells unnecessarily. At the low
  mass end, our refinement scheme results in an increased number of smaller
  mass cells, as expected. This drops off at the minimum cell mass, which we
  denote with a vertical red line. Similarly, in the central panel we show the
  distribution of stellar particle masses. Here, the distribution is the same
  for the case with refinement as that without refinement. We suppress star
  formation in the refinement region, which prevents smaller star particles
  from being formed, which could cause unwanted numerical heating. In the
  right-hand panel, we show the time evolution of star formation rates with
  and without our refinement scheme. The star formation rates are broadly the
  same regardless of the refinement scheme being used or not, with the
  slightly larger star formation rates at late times in the refinement case
  caused by the slightly lower rate of accretion onto the central black hole
  (see Figure~\ref{bh_ref}).} 
\label{part_dist}
\end{figure*}

In Figure~\ref{part_dist} we validate our super-Lagrangian refinement scheme
on the models of isolated disc galaxies which are evolved for $10\, {\rm Gyrs}$  in total. In the
left-hand panel we plot the distribution of cell masses normalized to the
target gas mass. With the standard de-/refinement of gas cells present in {\small
  AREPO} the cell mass distribution peaks around the target value and extends
within a factor of $\sim 2$ from the peak value (blue curve). With our super-Lagrangian
refinement around black holes the distribution of cell masses at the high mass
end (beyond the target value) is essentially identical. This validates our
choice of $R_\mathrm{cell}^\mathrm{max}=0.5\;h_\mathrm{BH}$ and explicitly
demonstrates that we are not unnecessarily de-/refining the cells once they
enter or leave the black hole refinement region. The distribution of cell masses
below the target value is however markedly different. This is caused by the
cells within our super-Lagrangian refinement region where we extend the
dynamical range in mass by more than six orders of magnitude. Note that the
low mass cutoff is due to the imposed minimum cell mass, as denoted with the
vertical dashed line.

In the central panel of Figure~\ref{part_dist} we show the mass distribution
of stellar particles with and without the refinement scheme. The two show very good
agreement, indicating that our refinement scheme is not introducing a broad
range of stellar particle masses, which could lead to unwanted \textit{N}-body
heating. Furthermore, the right-hand panel of Figure~\ref{part_dist} demonstrates
that the star formation rate evolution is not artificially affected in
simulations with super-Lagrangian refinement where we do not allow gas cells 
in the refinement region to be star forming. In fact the star formation rates are
broadly the same regardless of the refinement scheme being used or not, with
the slightly larger star formation rates at late times in the refinement case
caused by the slightly lower rate of accretion on to the central black hole. 

\begin{figure}
 \includegraphics[width=0.99\columnwidth]{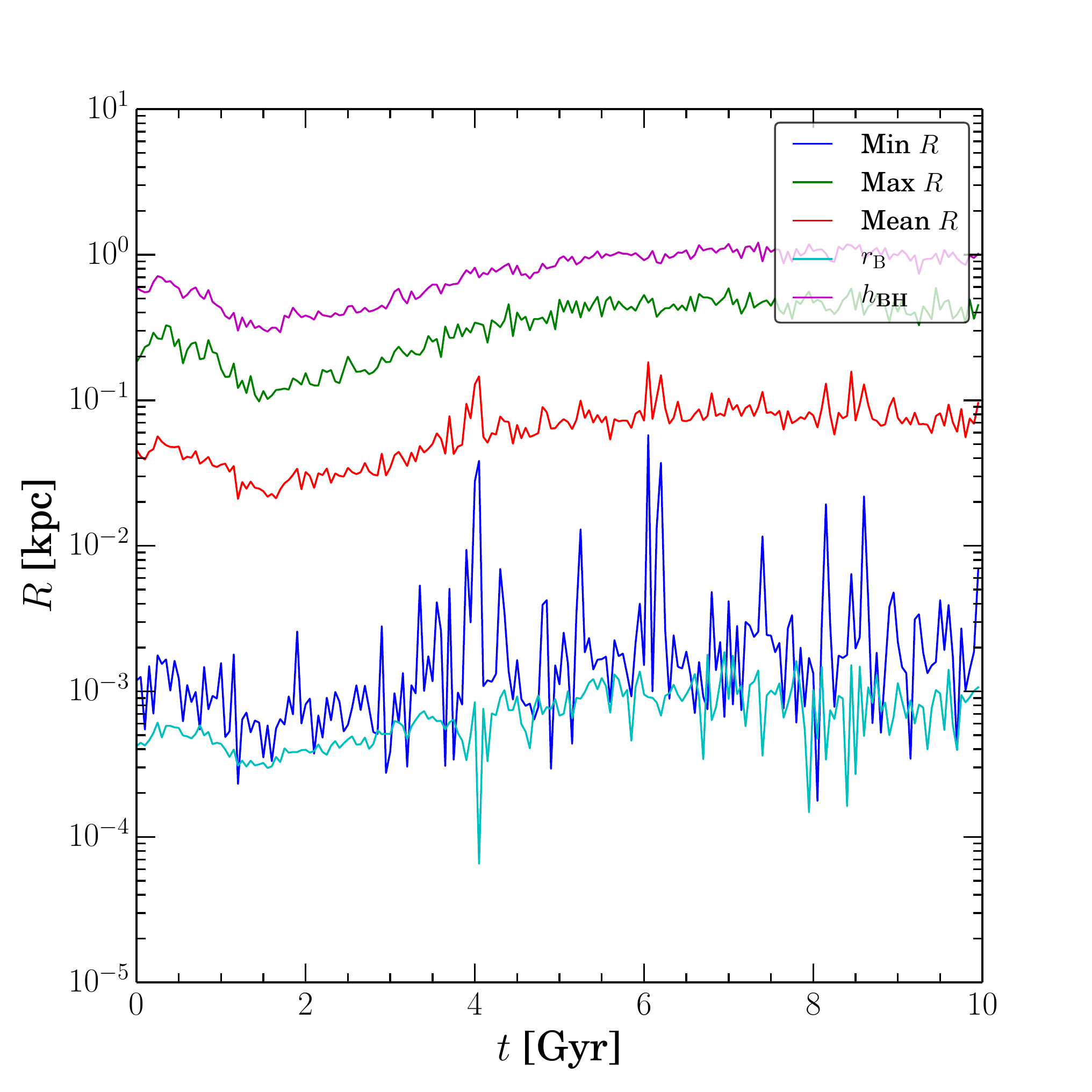}
 \caption{The distribution of cell radii within the refinement region. Here we
   show the minimum (blue), the maximum (green) and the mean (red) cell radii
   for cells within the refinement region as a function of time. We also show
   $r_\mathrm{B}$ as a function of time (cyan). Our refinement parameters
   here are $R_\mathrm{ref} = h_\mathrm{BH}$, $R_\mathrm{cell}^\mathrm{max} =
   0.5h_\mathrm{BH}$ and $R_\mathrm{cell}^\mathrm{min} = r_\mathrm{B}$, where
   $h_\mathrm{BH}$ is the black hole smoothing length (purple).}
 \label{bondi_comp}
\end{figure}

Figure~\ref{bondi_comp} shows more explicitly how our refinement scheme
affects the gas cell size. Here, we plot the minimum (blue), maximum (green)
and mean (red) of the cell radii within the black hole smoothing length,
$h_\mathrm{BH}$, which we also show (purple). We compare these to the
instantaneous Bondi radius, $r_{\rm B}$, (cyan)
over the whole duration of the run, i.e. $t = 10 \, {\rm Gyrs}$. With our choice of 
parameters the refinement region spans $0.5 - 1\, {\rm kpc}$ in radius, the
mean cell size within the refinement region is of the order of $0.05 - 0.1\, {\rm kpc}$,
while the minimum cell radius $ r_\mathrm{min} \le 2 r_{\rm B}$ for the majority of
the simulation timespan. This implies that we are not only probing the spatial
regions very close to the Bondi radius, but that we are also resolving much
better the gas structure in its surroundings which helps to estimate more
accurately the gas density and sound speed that are used in
equation~\ref{bondi_rate_eq}.

\begin{figure*}
\includegraphics[width=0.32\textwidth]{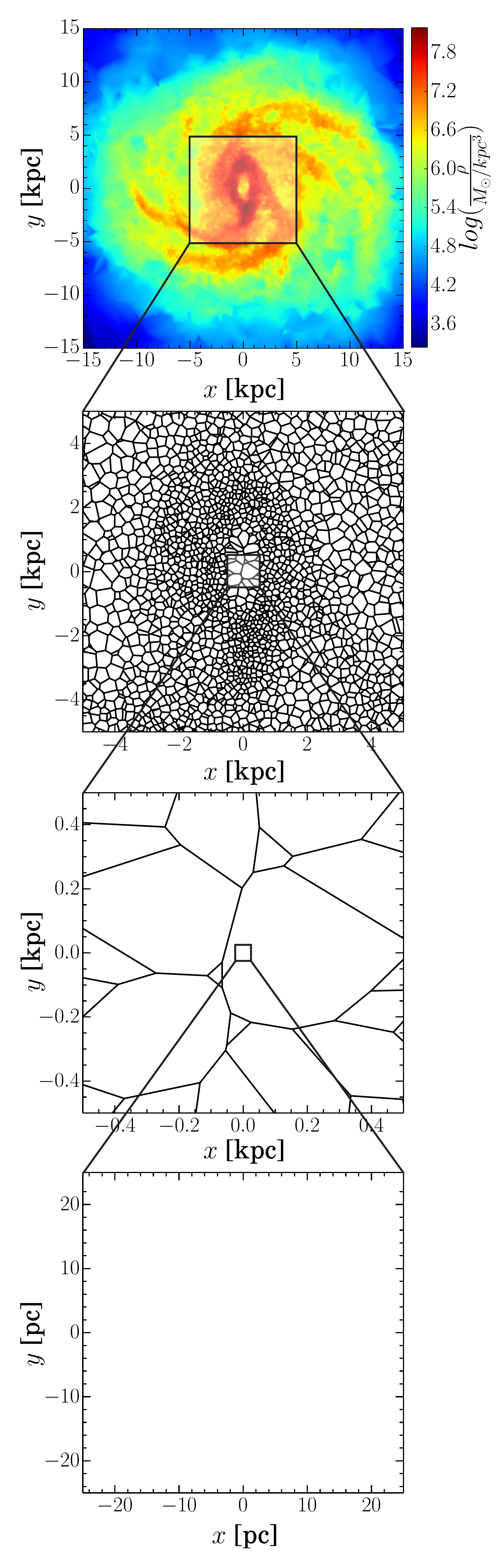}
\includegraphics[width=0.32\textwidth]{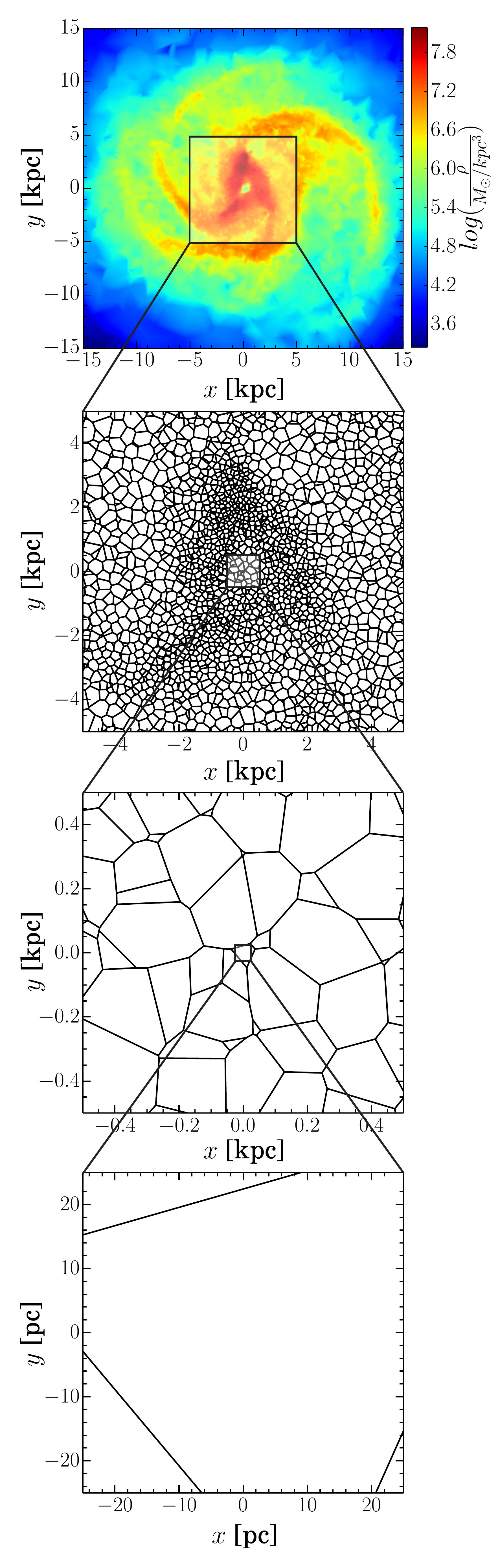}
\includegraphics[width=0.32\textwidth]{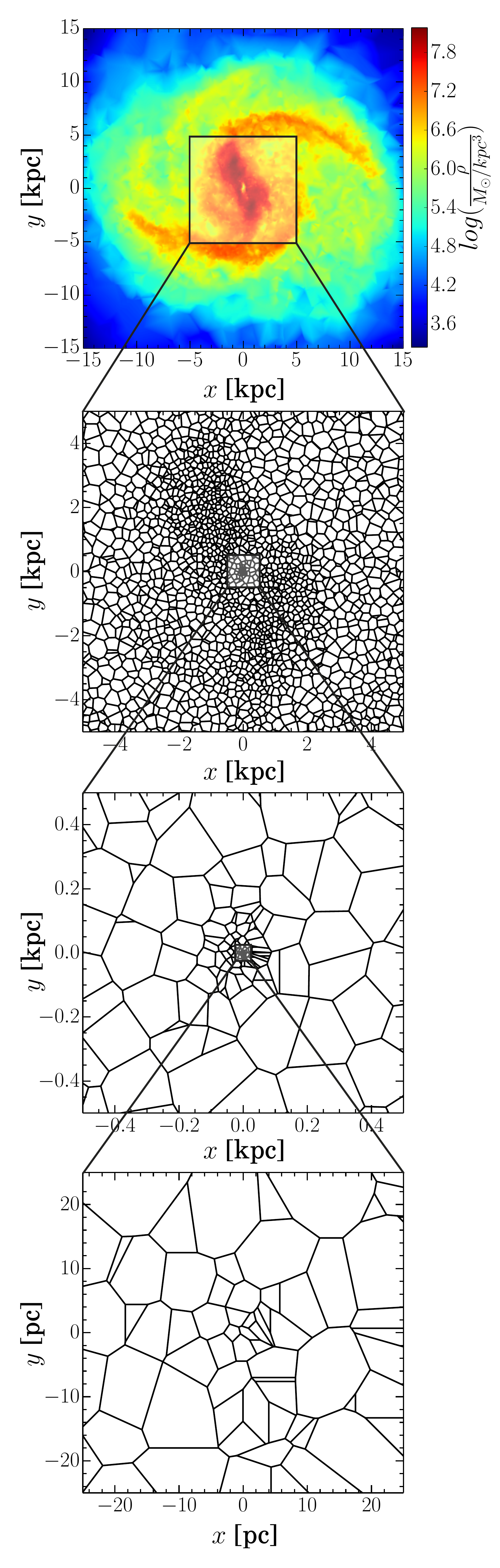}
\label{fg:aggref}
  \caption{Illustrations of our refinement method. All images show the Voronoi
    mesh, with cells coloured according to the gas density on the large scale
    plots in the top panels. The top row shows slices throughout the $x-y$
    plane in the central 
    $30\, \mathrm{kpc}\times30\, \mathrm{kpc}$ region centred on the black hole. The
    bottom rows show a zoomed-in plots of the central $10\,
    \mathrm{kpc}\times10\, \mathrm{kpc}$, $1\, \mathrm{kpc}\times1\,
    \mathrm{kpc}$ and $50\, \mathrm{pc}\times50\, \mathrm{pc}$ region,
    respectively. The first column 
    shows the case for no 
    refinement, whilst the second column shows the same, but for a simulation including
    moderate refinement with parameters $R_\mathrm{ref} = h_\mathrm{BH}$,
    $R_\mathrm{cell}^\mathrm{max} = 0.5h_\mathrm{BH}$,
    $R_\mathrm{cell}^\mathrm{min} = 10r_\mathrm{B}$, with
    $M_\mathrm{min}=10^{2}M_{\mathrm \odot}$. Third column shows a simulation with more
    aggressive parameters $R_\mathrm{ref} = h_\mathrm{BH}$,
    $R_\mathrm{cell}^\mathrm{max} = 0.5h_\mathrm{BH}$,
    $R_\mathrm{cell}^\mathrm{min} = r_\mathrm{B}$, with
    $M_\mathrm{min}=10^{-2}M_{\mathrm \odot}$.}
 \label{ref_mesh}
\end{figure*}

\begin{figure*}
\centering
\vbox{
\hbox{\includegraphics[width=0.5\textwidth]{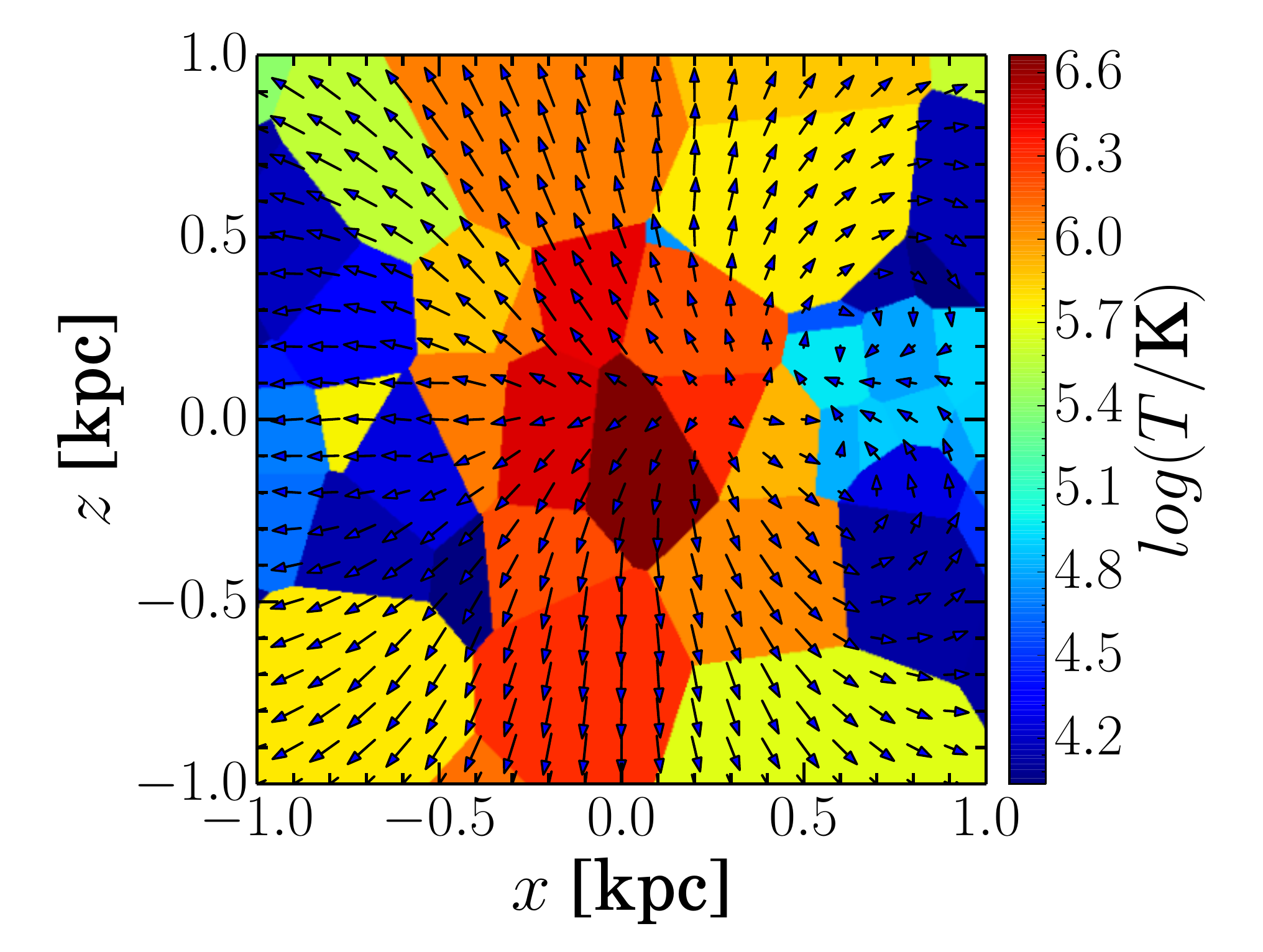}
 \includegraphics[width=0.5\textwidth]{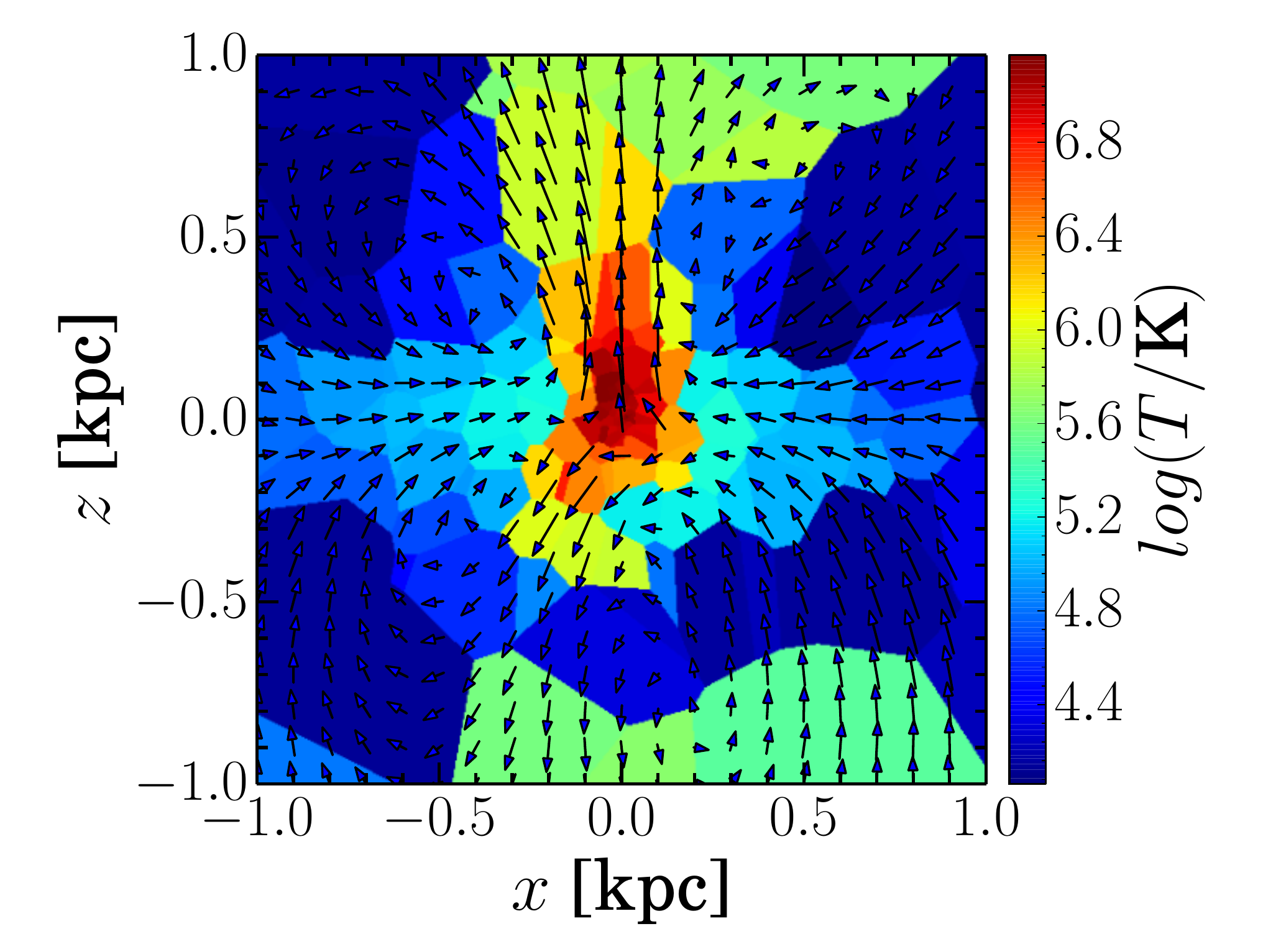}}
\hbox{\includegraphics[width=0.5\textwidth]{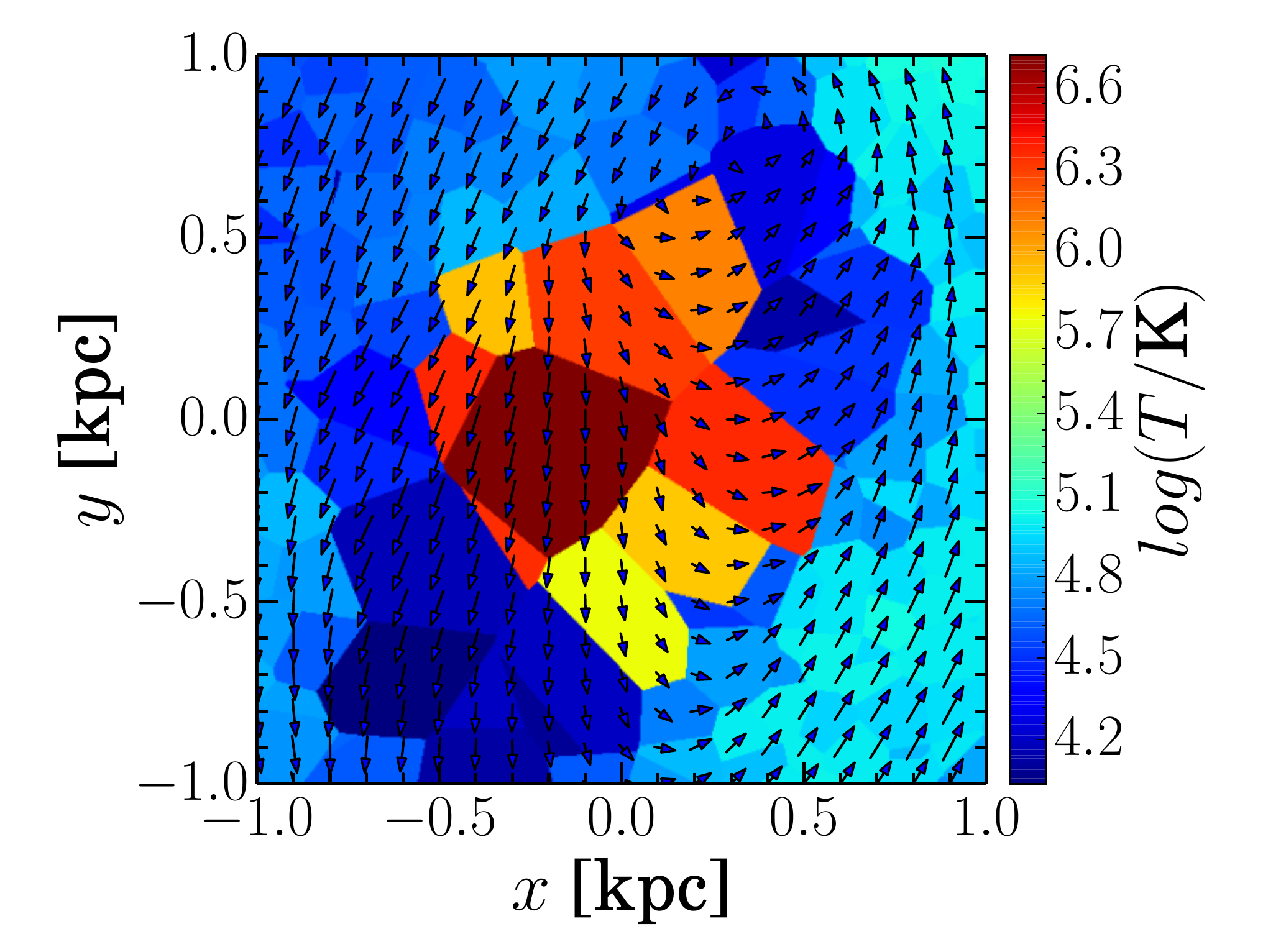}
 \includegraphics[width=0.5\textwidth]{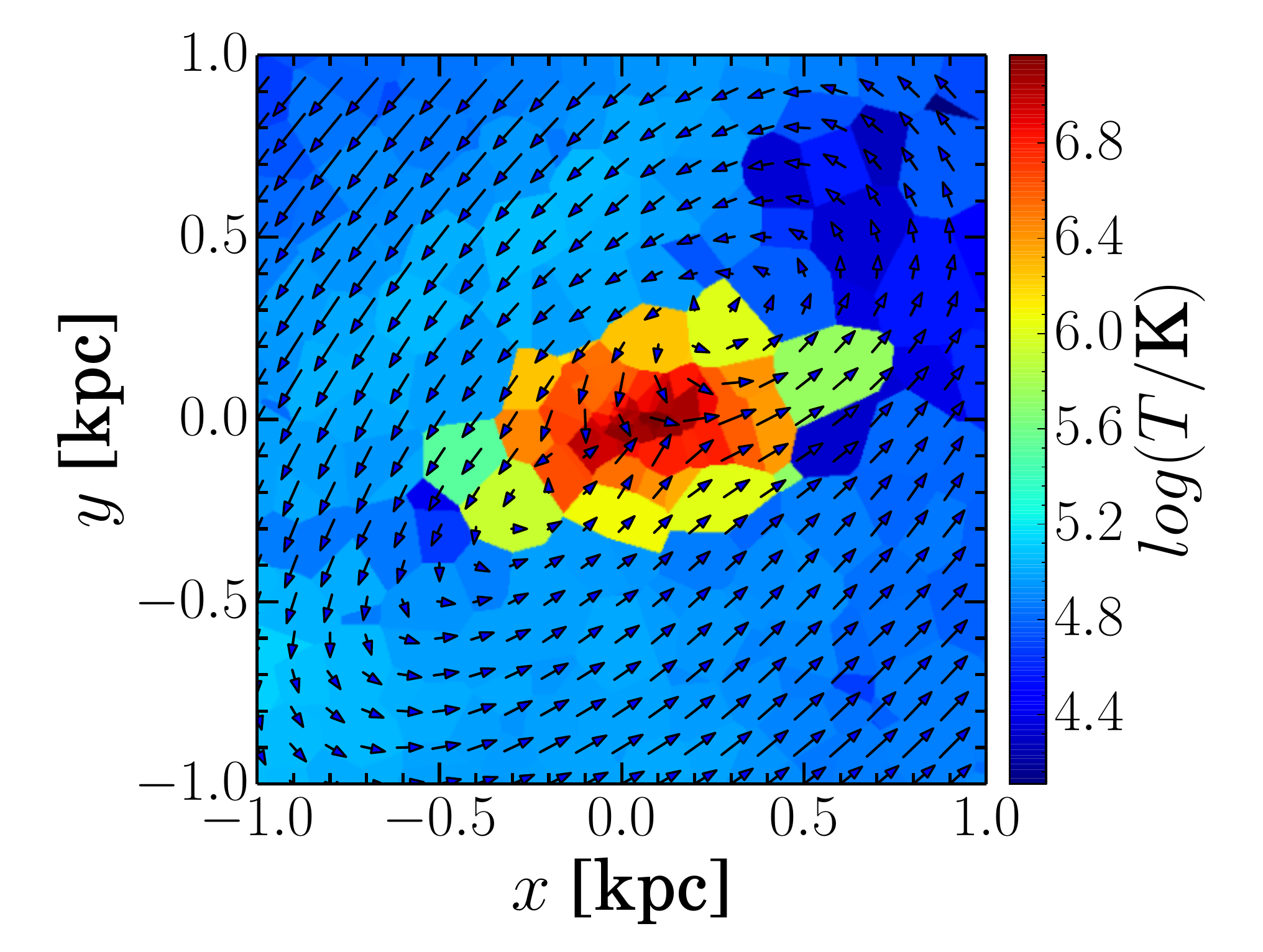}}
}
  \caption{\label{vmaps}Temperature maps of the gas in the central region for
    simulations without (left-hand panels) and with
    (right-hand panels) super-Lagrangian refinement using thermal feedback. Each
    Voronoi cell is coloured according to the mean gas temperature within the cell. Overplotted
    arrows indicate the direction and relative speed of the gas flow.
      The mean arrow size corresponds to $\sim 80 \, {\rm km \,s}^{-1}$ while the largest
      arrows are for $\sim200 \, {\rm km \,s}^{-1}$ in the left-hand panels, and
      $\sim300 \, {\rm km \,s}^{-1}$ in the right-hand panels. In the
    top row, we show a slice through the $x-z$ plane centred on the black
    hole (edge-on view) at a time of $1.5$ Gyr. Here, we see the effect 
    of feedback on the velocity of the gas,
    causing it to buoyantly rise perpendicular to the disc before
    settling back down at larger radii. In the non-refined simulation, the
    feedback heated gas dominates the central region whilst, in the refined
    simulation, the cold gas component that forms the outer accretion disc is
    clearly evident. This can also be seen in the bottom row, where we plot a
    slice through the $x-y$ plane centred on the black hole (face-on view) at 
    $1.0$ Gyr, before the disc is significantly disrupted. Here, we can see
    how the rotation structure of the gas is affected by the presence of the
    feedback, with the non-refined case again suffering from resolution
    problems in the central most region.}
\end{figure*}

\subsection{The effect on gas properties and black hole growth}

Figure~\ref{ref_mesh} shows how our refinement scheme works in practice. In
the top panels we plot the large scale gas density distribution in a box of
$30\, \mathrm{kpc}$ on a side, while in rows 2, 3 and 4 we show the
Voronoi tessellation across the $x-y$ plane as we zoom-in towards the central
black hole. The left-hand panel shows the results for a standard simulation
when our black hole refinement scheme is turned off. As the black hole accretes matter, energy is injected into the surrounding gas cells, increasing their
temperature and decreasing their density. The code maintains approximately
constant mass cells, so they necessarily increase in volume. This
significantly coarsens the resolution around the central region. This is a numerical problem present in all simulation codes which maintain roughly constant mass cells (or particles, i.e. in SPH) and/or which do not refine regions around black holes specifically to cure this problem \citep[see also][]{Vogelsberger:13}. As such, the vast majority of simulations in the literature suffer from the same effect.

In the middle and right-hand panels, we show the same situation with our
refinement scheme turned on. Here, the lower density cells caused by the
feedback are still evident, but the region affected is much smaller. Indeed,
in the simulation with more aggressive refinement parameters (right-hand panels), while the large scale fluid properties are the same, the structure of the innermost gas is
very different - the edge of the hot bubble is resolved, and higher density
gas is able to reach closer to the black hole and accrete.

Figure~\ref{vmaps} shows the effect that our refinement technique has on the central velocity and temperature structure. We show the temperature of the Voronoi mesh cells present in a slice in the region around the black hole for simulations using the thermal feedback prescription. The arrows
overlaying each plot show the velocity field of the gas cells, interpolated
over the slice. The top row shows slices through the $x-z$ plane (edge-on view). Here, we see
the effect of the thermal feedback in heating bubbles of gas that are forced to rise
buoyantly out of the disc. This is to be contrasted with the cold disc present
in the $z=0$ plane, which is an extension of the gaseous disc on large scales
and which provides the dense gas responsible for feeding the black hole. There is a stark difference over the central region - in the non refined case the feedback
overwhelms the central region, with cells ballooning out of the disc. In
particular, much of the structure of the outflow in the region closest to the
black hole is unresolved. This can be contrasted with the refinement
simulation, where the cold gaseous disc is significant down to scales much closer to the black hole. The feedback driven outflow is also better
resolved. The bottom row shows similar plots for the $x-y$ plane (face-on view). In both
cases, there is evidence of rotation and a centrifugally supported disc. We can see that in the refinement case, however, the disc is better resolved,
and the refinement prevents the feedback from destroying this structure. In addition, the central low density bubble is much smaller in the refinement case.

\begin{figure}
 \centering
 \includegraphics[width=0.45\textwidth]{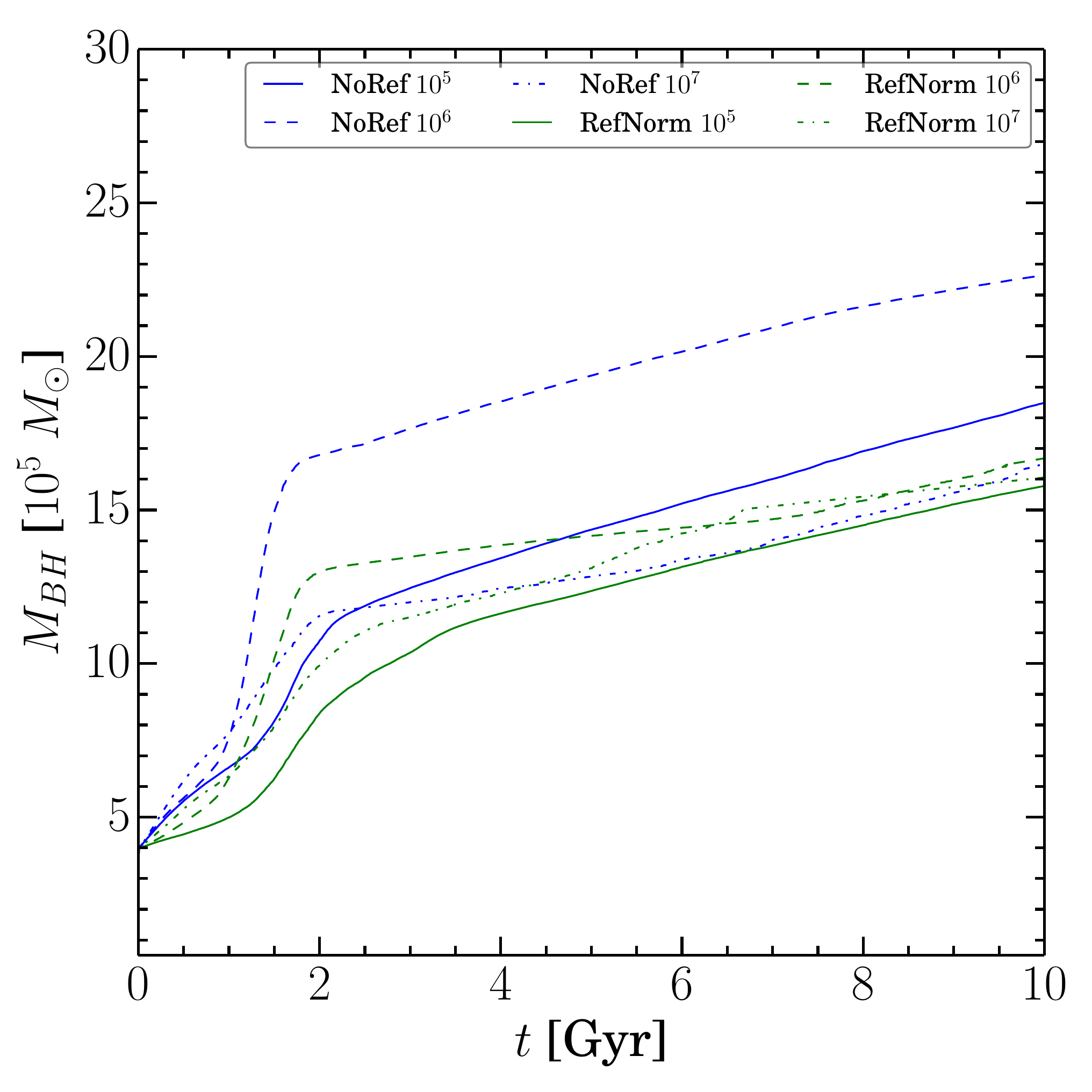}
  \caption{The evolution of the black hole mass as a function of time for
    simulations with no refinement (blue) and with refinement (green) for
    three different resolutions ($10^5$, $10^6$ and $10^7$ initial gas
    cells). All simulations here use the thermal feedback prescription. The
    convergence properties of super-Lagrangian refinement simulations are
    better, as expected.}
 \label{bh_ref}
\end{figure}

In Figure~\ref{bh_ref}, we show the effect of our refinement technique on the
mass growth rate of the black hole, for simulations with thermal feedback. The
simulations with refinement show somewhat lower rates of gas accretion than
those without. This can be explained in terms of the scheme's impact on the
estimated fluid parameters for the region surrounding the black hole. There
are two clear effects. First, the density of the fluid in the region around
the black hole, outside of the very centre, is higher. This is expected and is
because by increasing the local resolution, we are able to resolve higher
densities of accreting gas rather than smearing the same mass out over larger
cell sizes. However, the second effect is a consequence of depositing the
feedback energy into smaller cells. These rise to a higher temperature,
meaning that the sound speed is always above that for the non refined
case. This effect leads to a lowering of the density in the very centre of the
simulation and a subsequent suppression of the accretion rate. The change is
however not exceptionally large: this is because in this basic implementation
of feedback we are unable to take advantage of all of the improved resolution
that refinement gives. In addition to these differences, the simulations with
refinement show better convergence rate than those without. This is encouraging and, agreeing with the results of our Bondi inflow simulations, suggests that our refinement technique will allow future cosmological simulations to have better convergence properties, which is vital if they are to have higher predictive power.

\subsection{Changing the Feedback}

\subsubsection{The role of feedback}

To understand exactly how our refinement scheme affects the structure of the
gas, as well as how the choice of different feedback routines combine with
this, we need to isolate the effect of the injection of the feedback itself
from the subsequent coupled effect that this has on the sound speed and
density of the gas and, as a result, on the accretion rate itself. To this
end, we run a set of simulations of our isolated galaxy models, but with the
black hole accreting at a fixed, constant accretion rate. It is important that
this rate is not too high, or the surrounding gas will be blown away on a short time-scale, or too low, which will result in the different feedback algorithms being indistinguishable. We find that a value of $\dot{M}_\mathrm{BH} = 0.1 \dot{M}_\mathrm{Edd}$ (for the initial mass of the black hole) allows for this balance. We compare the impact of the feedback on the surrounding gas when using either the thermal or the momentum injection algorithms described above, and also how this changes when the feedback is injected over a duty cycle. We also investigate how using our new bipolar model of feedback changes the gas parameters.

\subsubsection{The effect on gas properties}

\begin{figure*}
\centering
\includegraphics[width=0.45\textwidth]{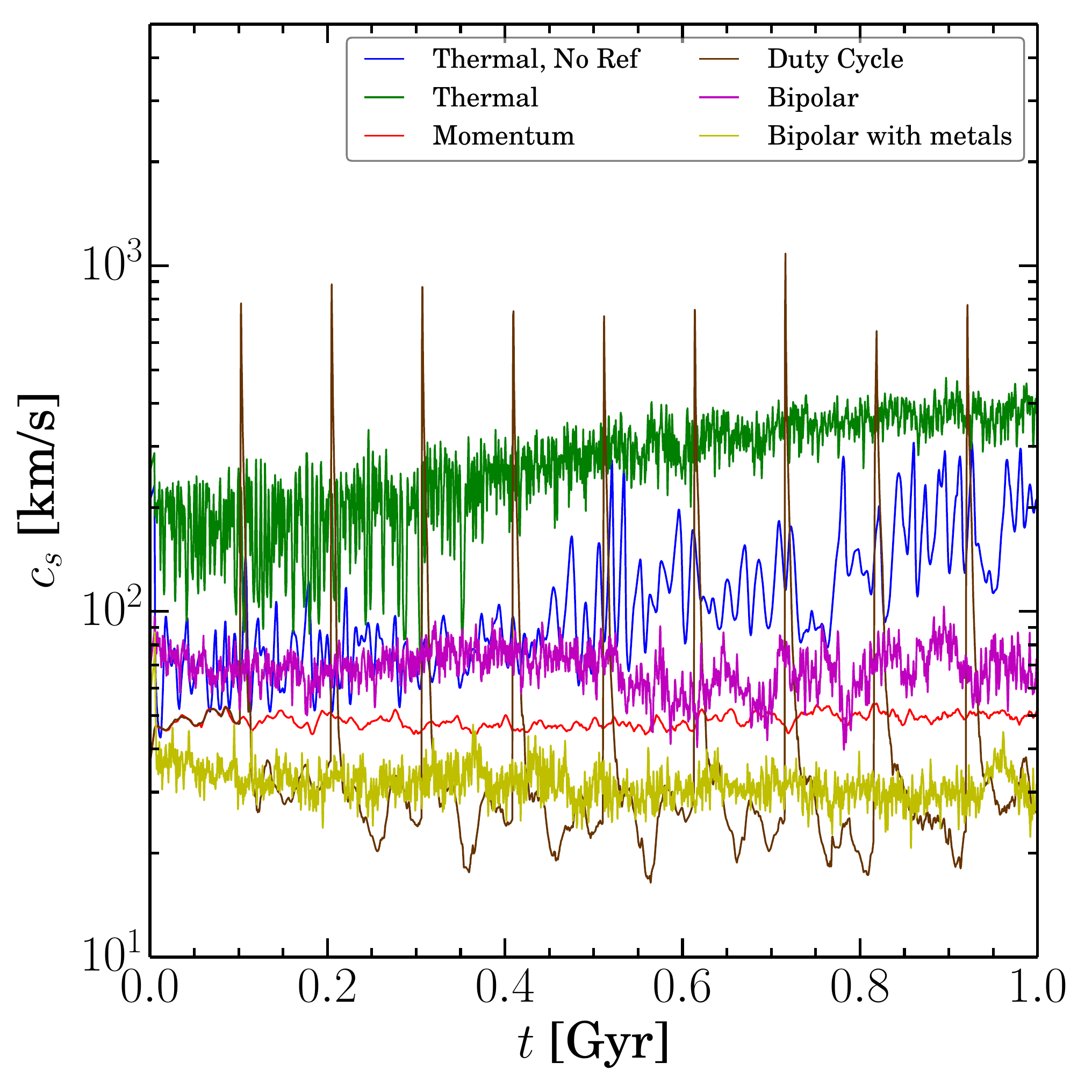}
\includegraphics[width=0.45\textwidth]{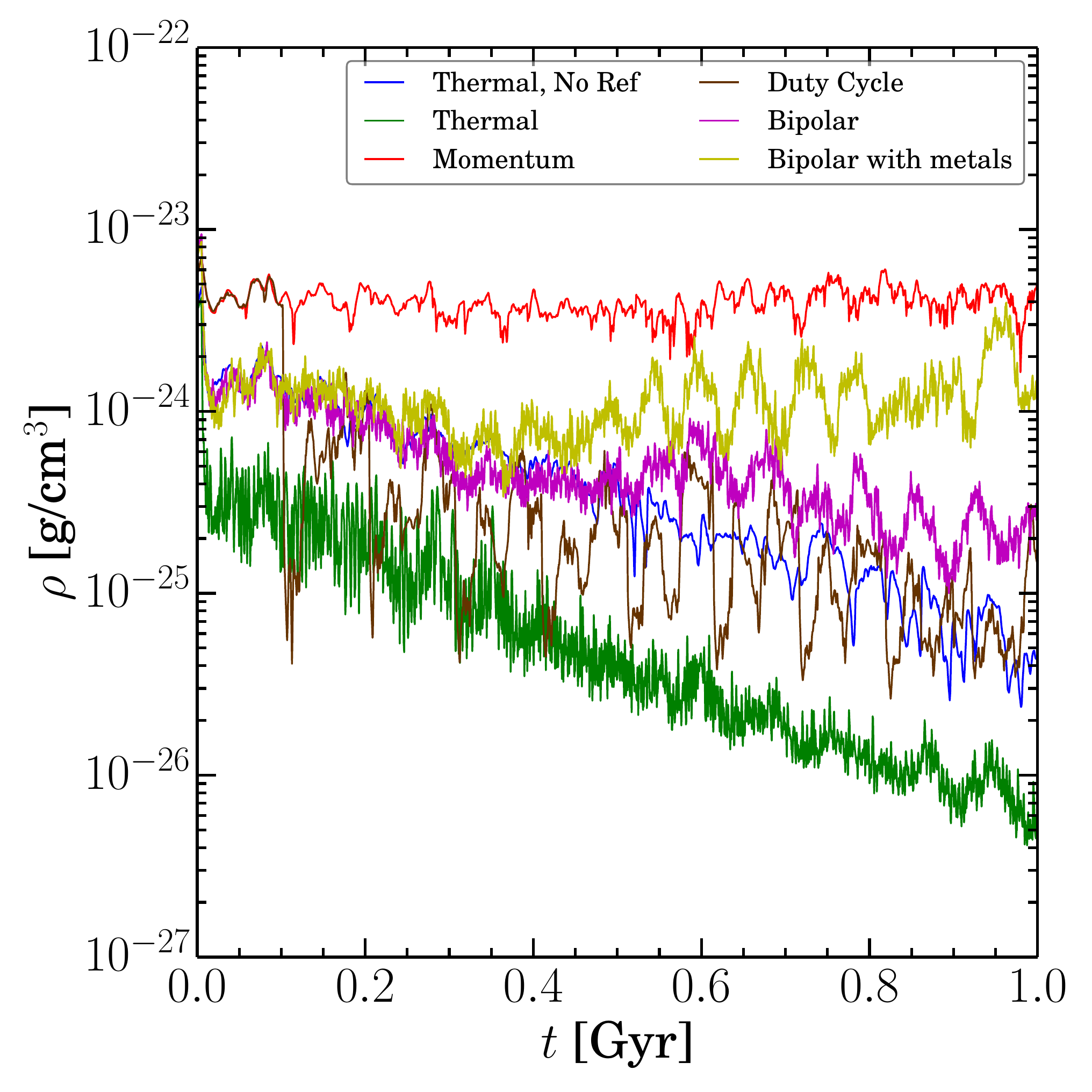}
\caption{The effect of the feedback algorithm on the sound speed and density
  of the gas. In the left-hand panel, we show the evolution of the sound speed
  of the gas and in the right-hand panel, we show the same for the density of the gas. In both cases, the value shown is that used in our estimation of the accretion rate, i.e. averaged over the smoothing length of the black hole. In blue, we show the case for thermal feedback with no refinement and in green that for thermal feedback but with refinement. We show the case for momentum feedback in red and in brown we show that for duty cycle feedback. We also show the results for our non isotropic model, both with (gold) and without (purple) metal line cooling enabled.}
 \label{cs_comp}
\end{figure*}

The most direct impact that the choice of feedback algorithm has is on the
temperature of the gas and, subsequently, the sound speed. We find that the
difference between the largest (simple, isotropic thermal feedback) and the
lowest (bipolar feedback with metal line cooling, or thermal feedback injected
in a duty cycle fashion) is consistently around a factor of 10 as shown in the
left-hand panel of Figure~\ref{cs_comp}. The green and blue lines show the sound speed for simple, isotropic thermal feedback for simulations with and without our refinement technique, respectively. Here, the masses of the cells in the centre of the refinement simulation are much smaller. As such, although the amount of energy distributed is the same, the peak temperature reached in these cells is much higher which, even when averaged over the smoothing length of the black hole, results in a higher sound speed. When momentum feedback (the red line) is used, the result is different - the gas is shock heated by secondary processes when the accelerated gas collides with accreting material. This results in a consistently much smaller sound speed. Indeed, we find that when using momentum feedback the black hole accretion is eventually shut off by a burst of feedback sufficient to dramatically reduce the density of the gas in the centre of the galaxy, rather than by increasing the sound speed (which dampens the accretion rate). Unsurprisingly, our simulations using bipolar feedback, both with and without metal line cooling, result in a consistently lower sound speed. Similarly, whilst the duty cycle run shows bursts aligned with each injection of feedback, the energy is quickly transported away and the overall sound speed is much lower than the standard thermal injection.

This raises a further point - that in addition to the rise in temperature of the gas, the subsequent accumulation or transportation of feedback energy is sensitive to the feedback routine. In the thermal feedback case, the sound speed of the gas close to the black hole steadily rises over the course of 1 Gyr, whilst for the other routines it is nearly constant. This implies a further compound effect that the feedback algorithm choice has in less controlled simulations.

\begin{figure*}
 \centering
 \includegraphics[width=0.45\textwidth]{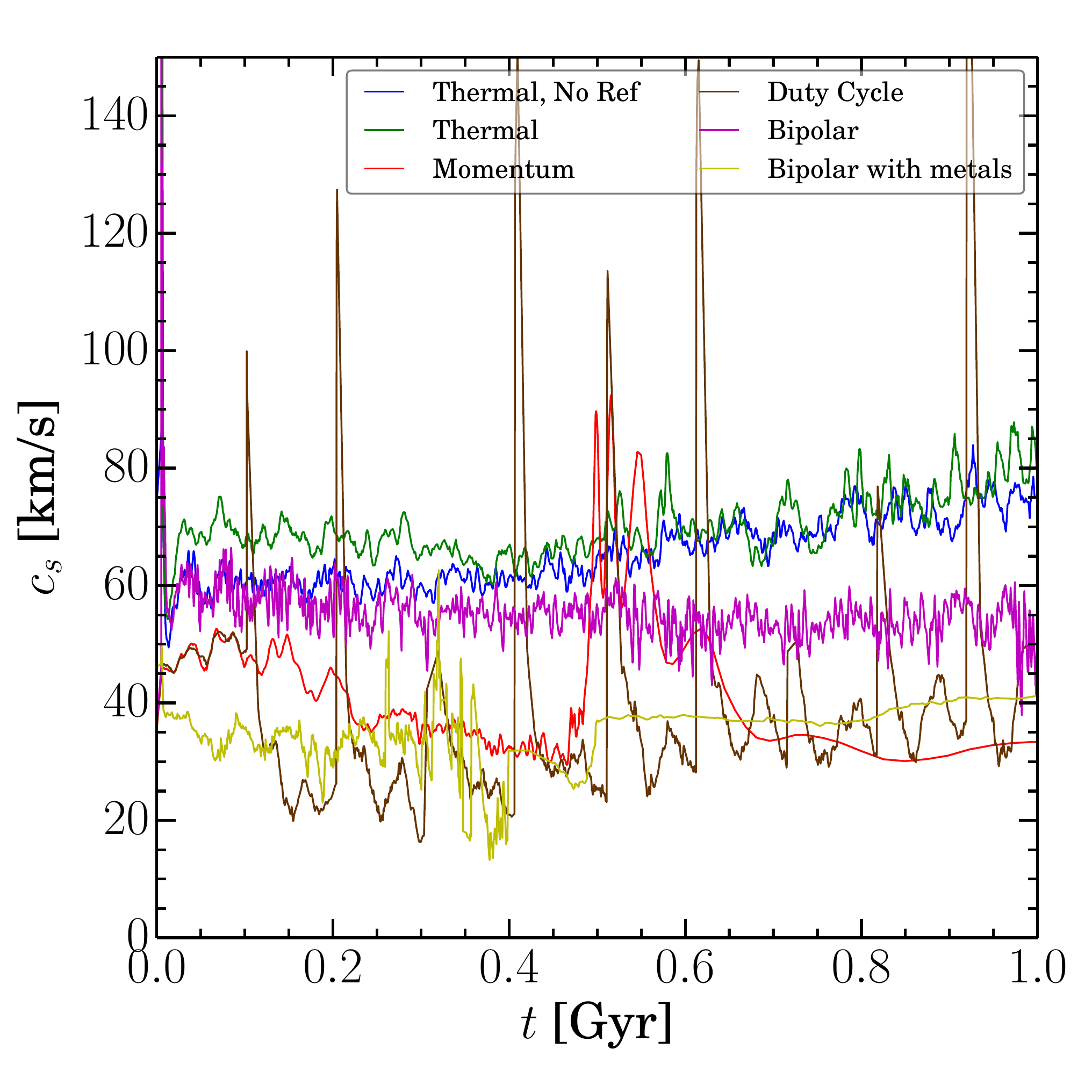}
 \includegraphics[width=0.45\textwidth]{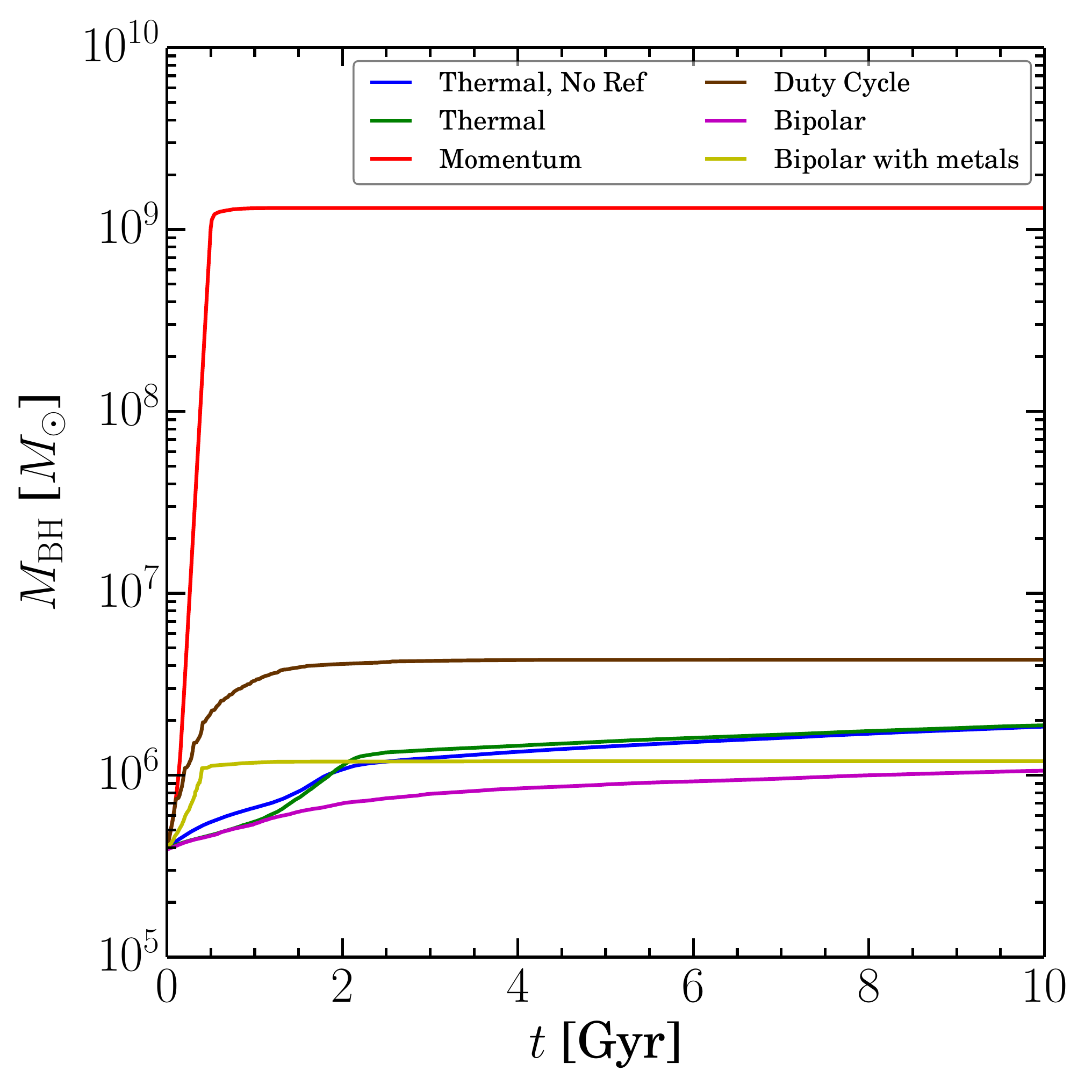}
 \caption{The effect of the feedback algorithm in self-consistent
   simulations. On the left, as in Figure~\ref{cs_comp}, we show how the sound
   speed evolves with time for different feedback routines, but now with
   self-consistent black hole accretion rate prescription. On the right we
   show the corresponding evolution of the black hole mass over the full
   $10 \, {\rm Gyrs}$ of time evolution.}
 \label{cs_not_fixed}
\end{figure*}

It might be argued that, in practice, this will not affect the average black
hole accretion rate over the relevant time-scales of galaxy formation because the accretion will enter a self-regulation phase - more powerful feedback will shut off the accretion on to the black hole, removing the energy for the feedback. We discuss the black hole accretion itself in the section below, however it is worth noting here that, even if this is the case for the commonly used Bondi accretion rate, when we consider the impact of gas angular momentum then the sound speed itself will take on a new importance. The magnitude of the effect of angular momentum is likely to be highly dependent on the sound speed of the gas: at high sound speeds, this effect is likely to be substantially decreased. We will discuss this in further detail in our next paper. Furthermore, when using our bipolar model, feedback has only an indirect effect on the accreting material, which results by construction in the simulation never quite achieving a steady self-regulation phase.

In the right-hand panel of Figure~\ref{cs_comp} we show the evolution of the
density of the gas surrounding the black hole. The impact here is generally
similar to that on the sound speed, but with some important differences. The thermal feedback causes a steady decrease in the density of the gas around the black hole, both due to the increase in temperature but also because mass is transported away, as hot gas rises buoyantly out of the galaxy. The same is true for the duty cycle feedback. The momentum feedback has the smallest effect on the density, because the momentum kicks given to the surrounding gas (whose strength are tied to the black hole mass) are insufficient to overcome the accreting material. Both bipolar models show similar behaviour in maintaining a higher density for a longer period of time into the simulation, because the feedback is restricted to a cone and as such the accreting material does not fall off in density as quickly. When metal line cooling is included, this effect is magnified as the cold disc becomes denser throughout the simulation, offsetting the effects of the feedback. 

\begin{figure*}
 \centering
 \includegraphics[width=0.45\textwidth]{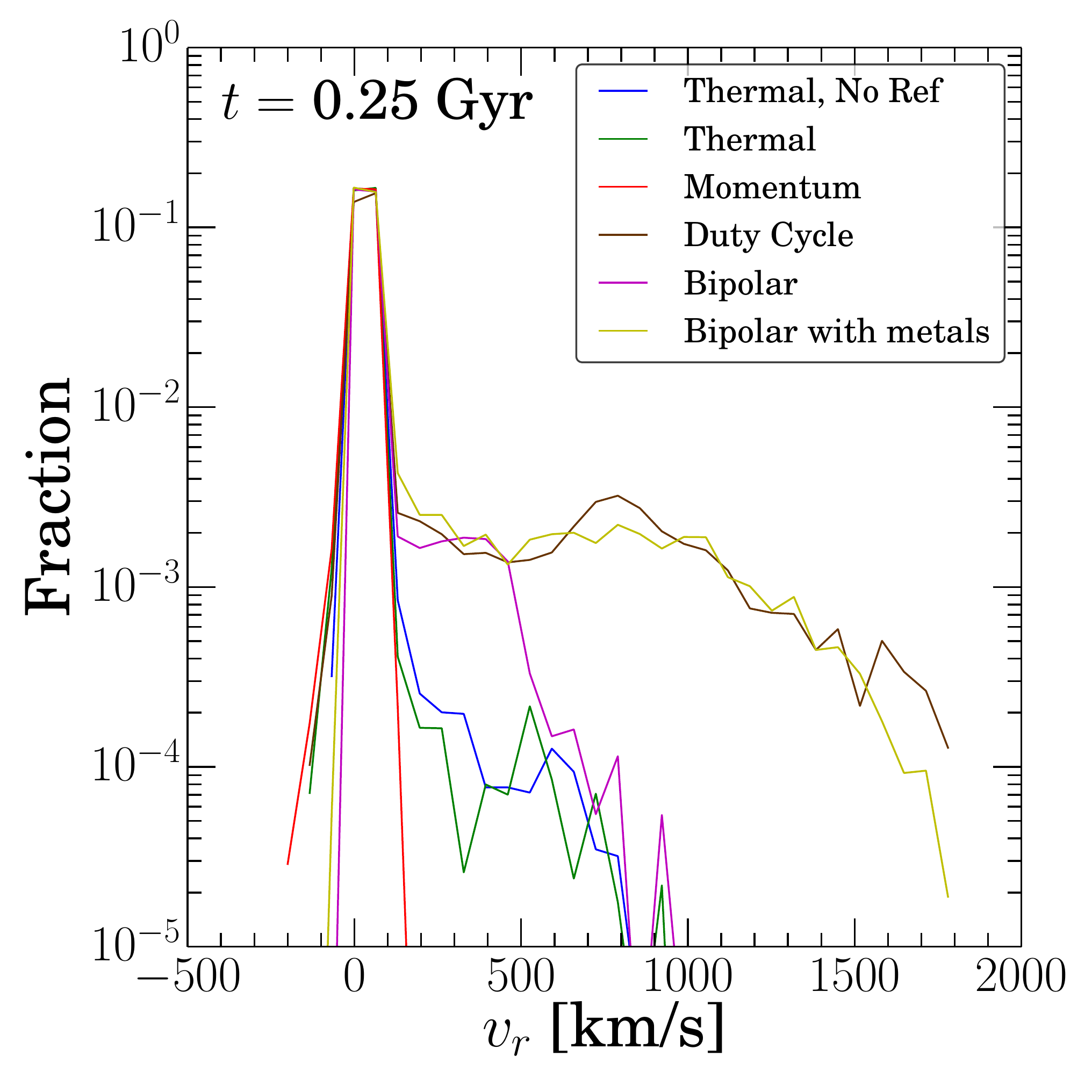}
 \includegraphics[width=0.45\textwidth]{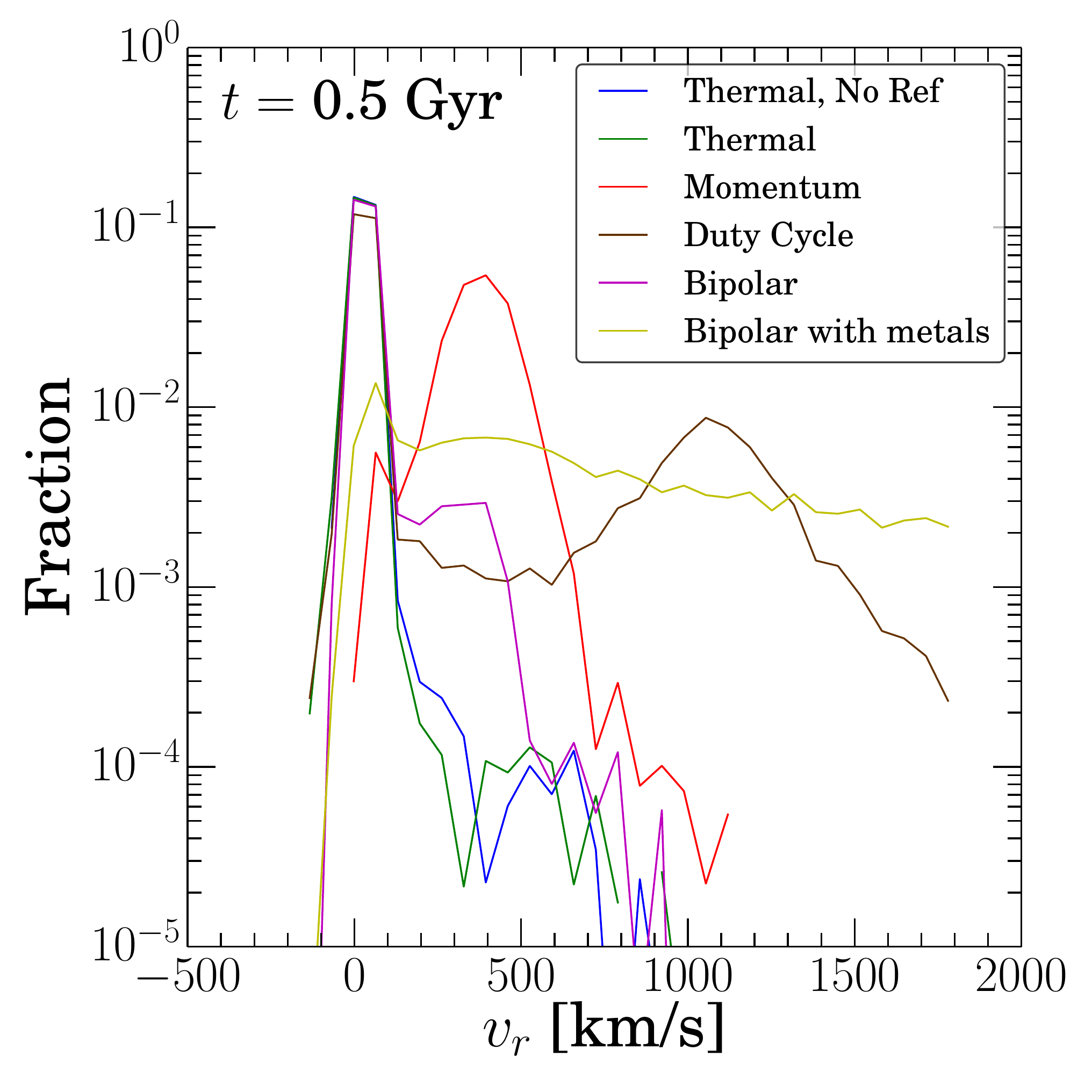}
 \caption{Radial velocity probability density functions for simulations with
   different types of feedback, with self-consistent black hole
   accretion. Positive velocities indicate material moving away from the black
   hole. Left-hand panel is for $t=0.25\, {\rm Gyr}$, while on the right
   $t=0.5 \, {\rm Gyr}$. At
   early times, there is no significant outflow for the thermal or momentum
   models, whilst the bipolar and the duty cycle models show high velocity
   outflowing material. At the later time, the black hole now has sufficient
   mass that in the momentum case (red) the feedback is able to drive a high
   velocity shell of material that can be clearly seen peaking around $400\,
   {\rm km \,s}^{-1}$. The differences is gas properties of simulated galaxies
   may provide a powerful method for constraining the feedback mechanism when these are compared to observed galaxies.}
 \label{vr_pdf}
\end{figure*}

Left-hand panel of Figure~\ref{cs_not_fixed} shows a similar plot of the sound
speed of the gas in the vicinity of the black hole, but this time for
simulations with self-consistent black hole accretion and feedback
rates. Here, the sound speed of the gas is on the whole much lower and the
differences between the different feedback regimes are not as pronounced. This
is the evidence of self-regulation - the overall amount of feedback is lower
because any increase in the sound speed shuts off accretion. Note that,
despite this, there are still notable differences in the sound speed - up to a
factor of 4, which as previously discussed are still very important in the
context of considering angular momentum. Furthermore, there can still be large
differences in the final mass of the black hole, even when the sounds speeds
are similar, as we will discuss next.

\subsubsection{The consequences for black hole growth}

In the right-hand panel of Figure~\ref{cs_not_fixed}, we look at how the
choice of feedback algorithm affects the black hole growth. The first thing to
note is that the algorithm itself has the largest single impact on the final
mass of the black hole. In particular, for our simulation with momentum
feedback, which does not directly heat the gas, the black hole grows at a high
accretion rate until the momentum feedback is powerful enough to drive a shell
out of the centre of the galaxy, which shuts off the accretion but only after
the black hole has grown by many orders of magnitude. In the duty cycle case, the final mass of the black hole is an order of magnitude higher than that in the isotropic, thermal case. This is largely the effect of the substantially smaller sound speed of the gas, which dominates over the density in the accretion rate, as well as the fact that when not Eddington limited the black hole grows as $\sim M_\mathrm{BH}^2$, so any early differences are magnified at later times. Perhaps equally important, however, is that if we do not assume isotropy and break the direct connection between the outflowing, feedback heated gas and the cold accreting gas that feeds the black hole, we weaken the ease with which the black hole can fall into a stable self-regulated regime. 

However, even though the black hole ends up at a similar mass for the
simulations with thermal and bipolar feedback, the growth history and, as
noted before, the gas properties are markedly different. A further example of
this can be seen in Figure~\ref{vr_pdf}, in which we show the probability
density function of the radial velocity of gas in the same set of
simulations. In 
particular, this allows us to characterize the properties of the outflow for
different types of feedback that, in principle, may provide a very useful
diagnostics once compared to observations. In the left-hand panel we show the
radial velocity at $t=0.25\, {\rm Gyr}$ and on the right that at $t=0.5\, {\rm
  Gyr}$. In both
cases, the simple thermal feedback fails to drive a significant outflow. We
find that this does eventually emerge, but only after longer periods of
time. By contrast the run with bipolar feedback drives an outflow from early
times, an effect that is magnified when metal line cooling is included. In the
duty cycle case, too, there is a large proportion of fast outflowing material
at $t=0.25\, {\rm Gyr}$, whilst the outflow in the momentum feedback run is
minimal. This changes at $t=0.5\, {\rm Gyr}$ - at this point, the black hole has
sufficient mass to drive an outflow that removes gas from the centre of the
galaxy in a fast moving shell that can be clearly seen, peaking at around
$400\, {\rm km \, s}^{-1}$. The effect of this can be seen in the clear shut off of the
accretion rate in Figure~\ref{cs_not_fixed}. By comparing these figures, we
 also see that the accretion for the duty cycle run and bipolar run with
metal line cooling has all but dried up by $0.5\, {\rm Gyr}$ and that both show a large
amount of material in a corresponding outflow. In these cases, the discrete,
large injections of energy in the duty cycle case lead to an outflow bunched
around a peak velocity of around $1200\, {\rm km \,s}^{-1}$, whilst the continuous feedback of the bipolar outflow leads to an extended distribution across a wide velocity range. By looking at the outflowing mass as a function of radius, we can confirm this effect - for the bipolar simulation, mass is outflowing for a wide range of radii, whilst the duty cycle simulation only shows fast moving shells of mass ejected from the central region. This indicates that by investigating the fraction of galaxies with outflows at specified radii we may be able to constrain the nature of feedback.

\section{Conclusions} 
\label{Conclusions}
In this paper we have presented a new scheme for increasing the resolution
around black holes in a super-Lagrangian fashion in full
hydrodynamical simulations of galaxy formation. Our scheme adaptively targets
on-the-fly spatial regions around black 
holes where the resolution is increased progressively up to the desired
minimum spatial scales without introducing any unwanted boundary effects.

We have implemented our scheme in the moving mesh code {\small AREPO}. We 
demonstrated that our refinement technique is able to reproduce the
theoretical Bondi rate at lower resolutions than otherwise possible and that
we are able to match higher resolution runs with lower CPU costs. We have also
demonstrated the flexibility of our technique and that using more aggressive
refinement parameters leads to a closer agreement with the theoretical
predictions, at a cost of increased CPU resources. Our work opens the
possibility that in future we will be able to perform simulations of black
hole accretion by allowing black holes to swallow material directly using
  a sink particle-like routine in combination with a small-scale sub-grid model
  for accretion (without
the need to estimate the Bondi rate), a technique which has previously been
prone to high stochasticity and inaccuracies. 

We have studied the effects of our novel implementation in simulations of
isolated Milky Way-like galaxy models. We found that our refinement technique
is able to increase the resolving power around the central black hole by up to
seven orders of magnitude in mass resolution, reaching scales of order of the
Bondi radius for the whole duration of simulations of $10\, {\rm Gyrs}$. This
allowed us to  resolve more accurately the gas properties in the vicinity of
the black hole and thus to estimate the accretion rate more robustly. We
  stress however that that our accretion rate estimate still depends on the
  sub-grid model employed which is parameterized via Bondi-Hoyle-like accretion
  (see equation~\ref{bondi_rate_eq}).

Taking advantage of our novel refinement method we have implemented
several different mechanisms of injecting black hole feedback, including the
injection of mass, thermal energy and/or momentum in both isotropic and
non-isotropic distributions. We found that the choice of feedback algorithm
can have a very large effect both on the characteristics of the gas in the
central region of our simulated galaxies as well as on large scales, and,
consequently, on the growth history of the central black hole. Specifically,
we found that momentum-only feedback leads to a growth of overmassive black
holes before the gas can be efficiently expelled from the central region of
the galaxy \citep[see also][]{Costa:14}. Instead, feedback schemes that
incorporate at the same time mass, energy and momentum injection in conjuction
with cooling to low temperatures via metals seem most
promising in generating both realistic black hole masses and persistent
large-scale outflows with velocities up to $1500 \,{\rm km\, s}^{-1}$.   

The obvious next step for future work will be to look at the observational
characteristics of the different feedback mechanisms in full cosmological
simulations, which will be required to properly quantify the differences
outlined above in a realistic environment. As {\small ALMA} observations continue, and
as multiple integral field unit (IFU) surveys (e.g. {\small MaNGA}, {\small
  SAMI}, {\small CALIFA}, {\small Atlas3D}) release their data we will, in the
near future, have unprecedented detail of individual galaxies as well as much
larger statistical samples of galaxies. These data will
provide new constraints on black hole feedback processes. Computational models
of AGN feedback will have to continue to move beyond matching black hole
scaling relations and large scale galaxy properties towards reproducing
the detailed thermo-dynamical properties of galaxies, if we are to make use of
the constraining power of the new data available. Our super-Lagrangian
refinement method will allow us to more reliably track black hole accretion
and feedback in full cosmological simulations and thus to gain further
insight into the growth history of galaxies and supermassive black holes in
our Universe.

\section*{Acknowledgements}
We thank Martin Haehnelt, Ewald Puchwein, Chris Reynolds and Volker Springel
for their useful comments and advice. MC is supported by the Science and
Technology Facilities Council (STFC). 
This work was performed on the following: the COSMOS Shared Memory system at DAMTP,
University of Cambridge operated on behalf of the STFC DiRAC HPC
Facility - this equipment is funded by BIS National E-infrastructure capital
grant ST/J005673/1 and STFC grants ST/H008586/1, ST/K00333X/1; DiRAC Darwin
Supercomputer hosted by the University of Cambridge
High Performance Computing Service
(http://www.hpc.cam.ac.uk/), provided by Dell Inc.
using Strategic Research Infrastructure Funding from
the Higher Education Funding Council for England and
funding from the Science and Technology Facilities Council; DiRAC Complexity
system, operated by the University of Leicester IT Services,
which forms part of the STFC DiRAC HPC Facility
(www.dirac.ac.uk). This equipment is funded by BIS
National E-Infrastructure capital grant ST/K000373/1 and
STFC DiRAC Operations grant ST/K0003259/1. DiRAC is
part of the National E-Infrastructure.

\bibliographystyle{mn2e} 
\bibliography{references}

\appendix
\section{The black hole smoothing length}
In Section~\ref{sec:hsml} we described our procedure for calculating the
  fluid parameters in the vicinity of the black hole. In this appendix, we
  discuss the results of additional numerical tests we have carried out to
  demonstrate the robustness of this approach, which is commonly used in the
  literature. 

\begin{figure*}
\includegraphics[width=0.32\textwidth]{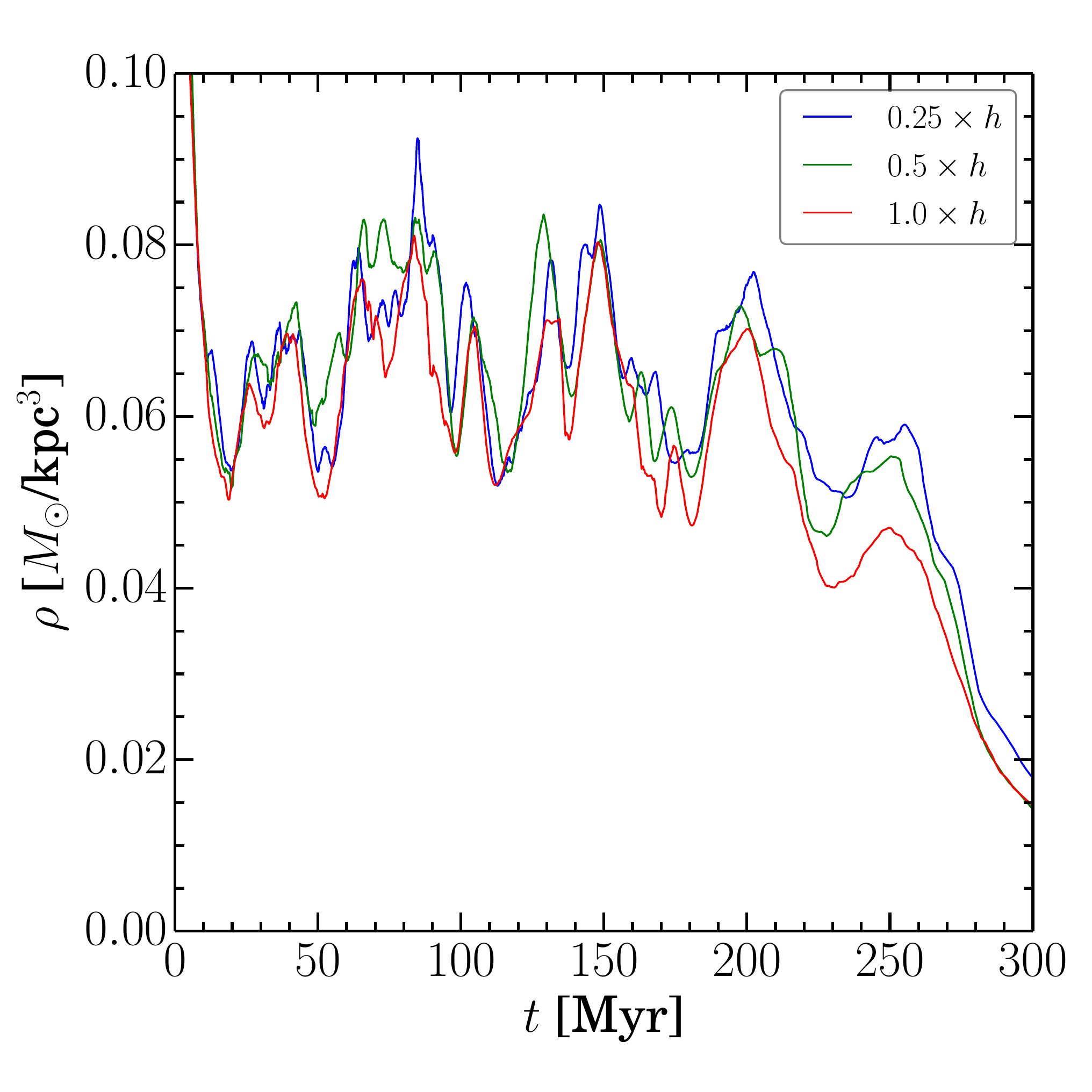}
\includegraphics[width=0.32\textwidth]{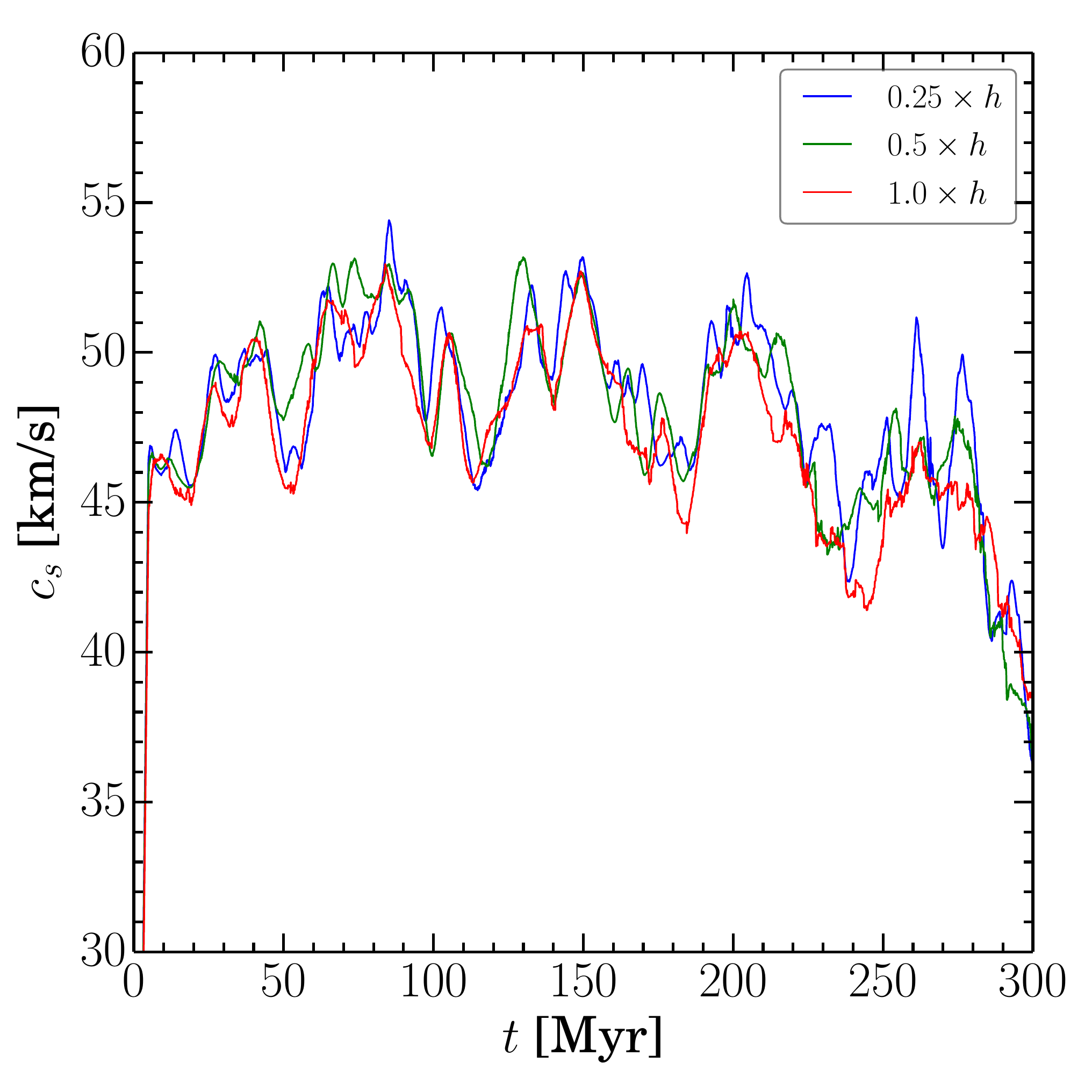}
\includegraphics[width=0.32\textwidth]{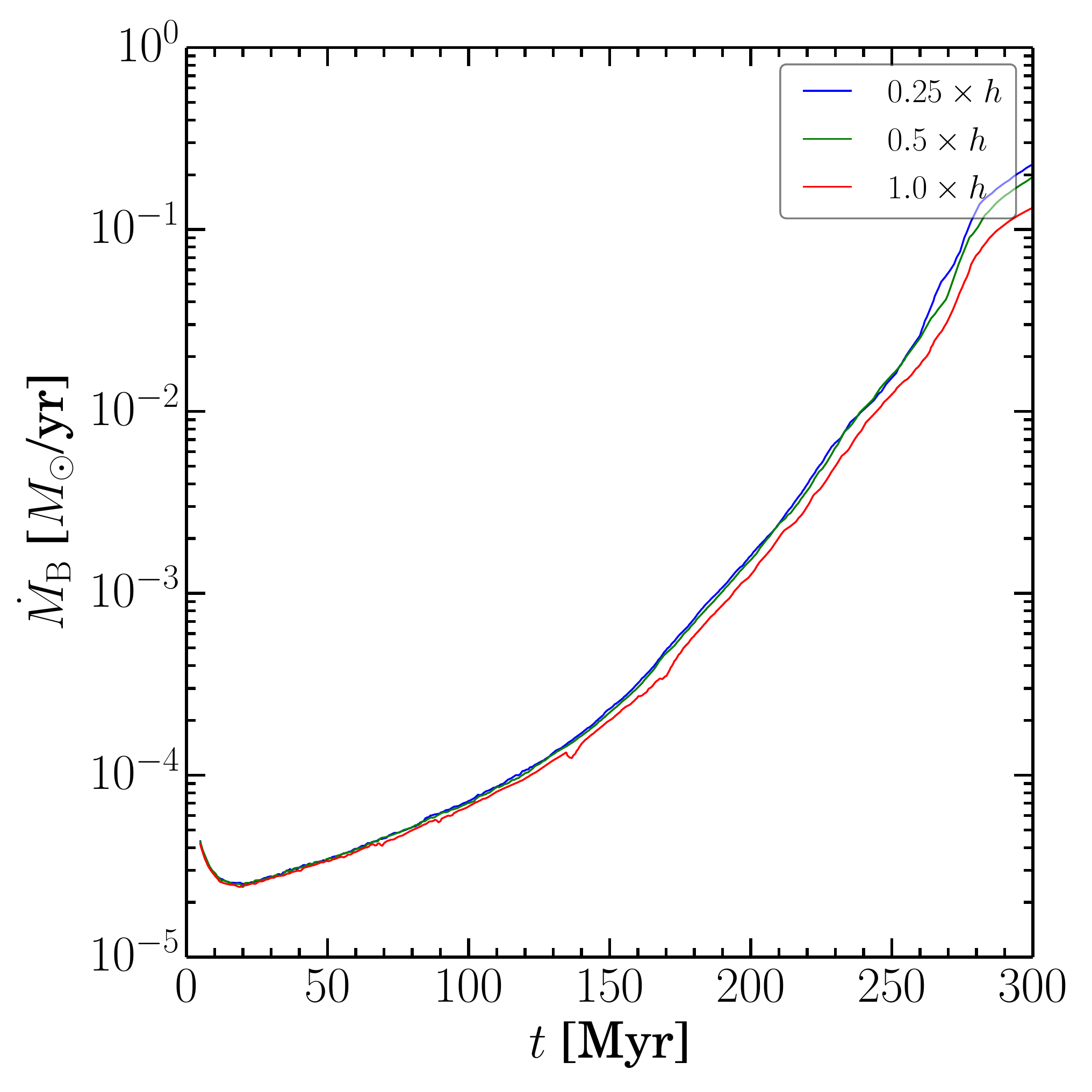}
  \caption{Fluid parameters and Bondi-Hoyle rate in simulations of an isolated
    disc galaxy ($10^5$ initial resolution elements) and no black hole
    feedback. All runs include our black hole refinement scheme. Changing the
    radius over which we calculate the weighted average of the fluid
    parameters does not substantially effect our results.}
 \label{hsml_low_comp}
\end{figure*}

To this end, we have run several simulations of our isolated disc galaxy
(using $10^5$ initial resolution elements) in which we vary the radius over
which weighted averages of the fluid parameters are estimated, for the
purposes of calculating the accretion rate. In all of these runs we enable
black hole refinement, using our standard parameters. To ensure that the tests
are consistent and to keep the tests as clean as possible we do not include
black hole feedback of any type in these simulations. In
Figure~\ref{hsml_low_comp} we show the results of doing this for three runs
where we estimate $\rho_\infty$ and $c_\infty$ at $1.0$, $0.5$ and $0.25$
times the black hole smoothing length. The results show that both the measured
quantities and the subsequently derived Bondi-Hoyle rate show good agreement
for the duration of the black hole growth phase.

In addition, we also show the same simulations carried out using $10^7$
initial resolution elements, with all simulations estimating the fluid
parameters at the full black hole smoothing length. These also show good agreement with each other, and good convergence with the comparable lower resolution run.

\begin{figure*}
\includegraphics[width=0.32\textwidth]{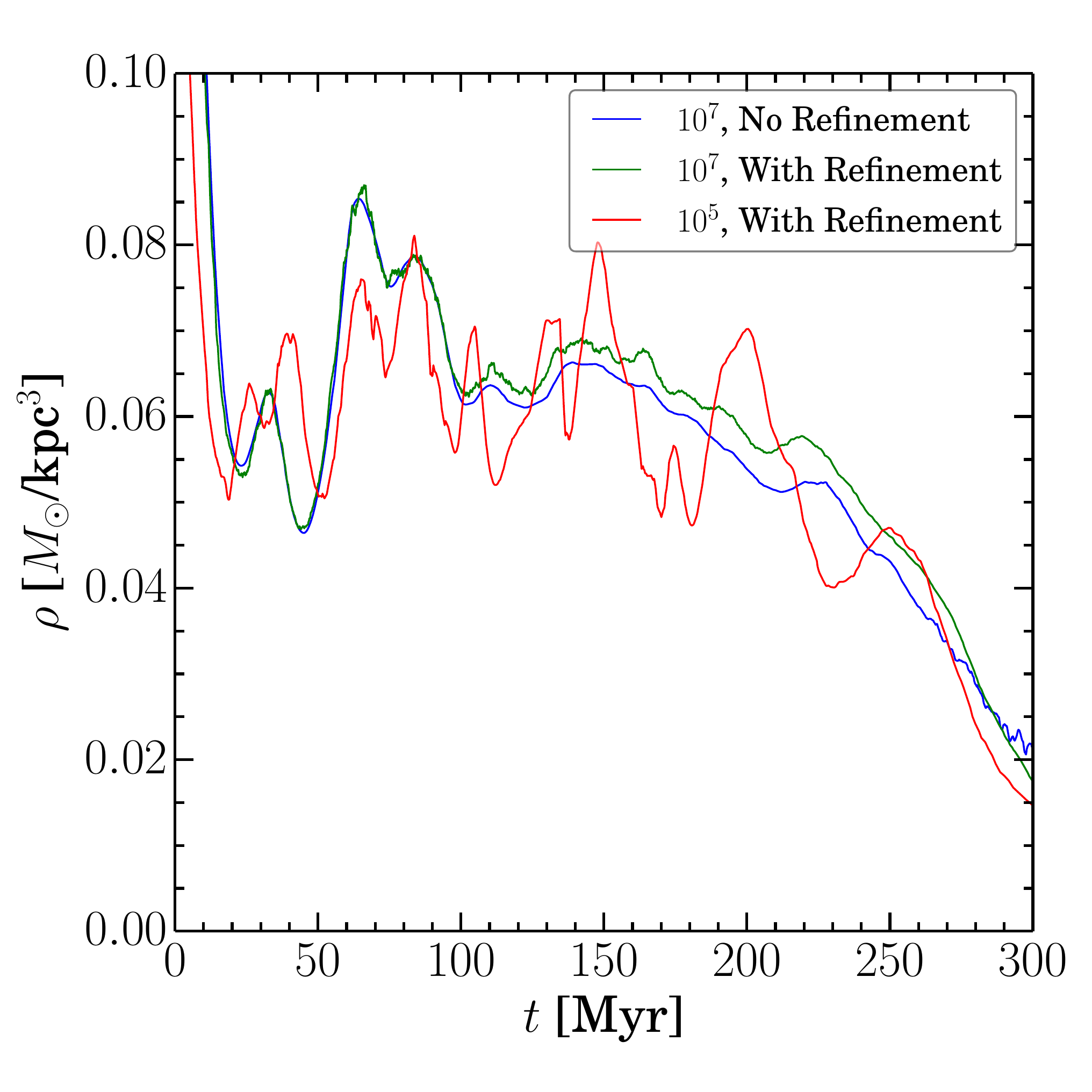}
\includegraphics[width=0.32\textwidth]{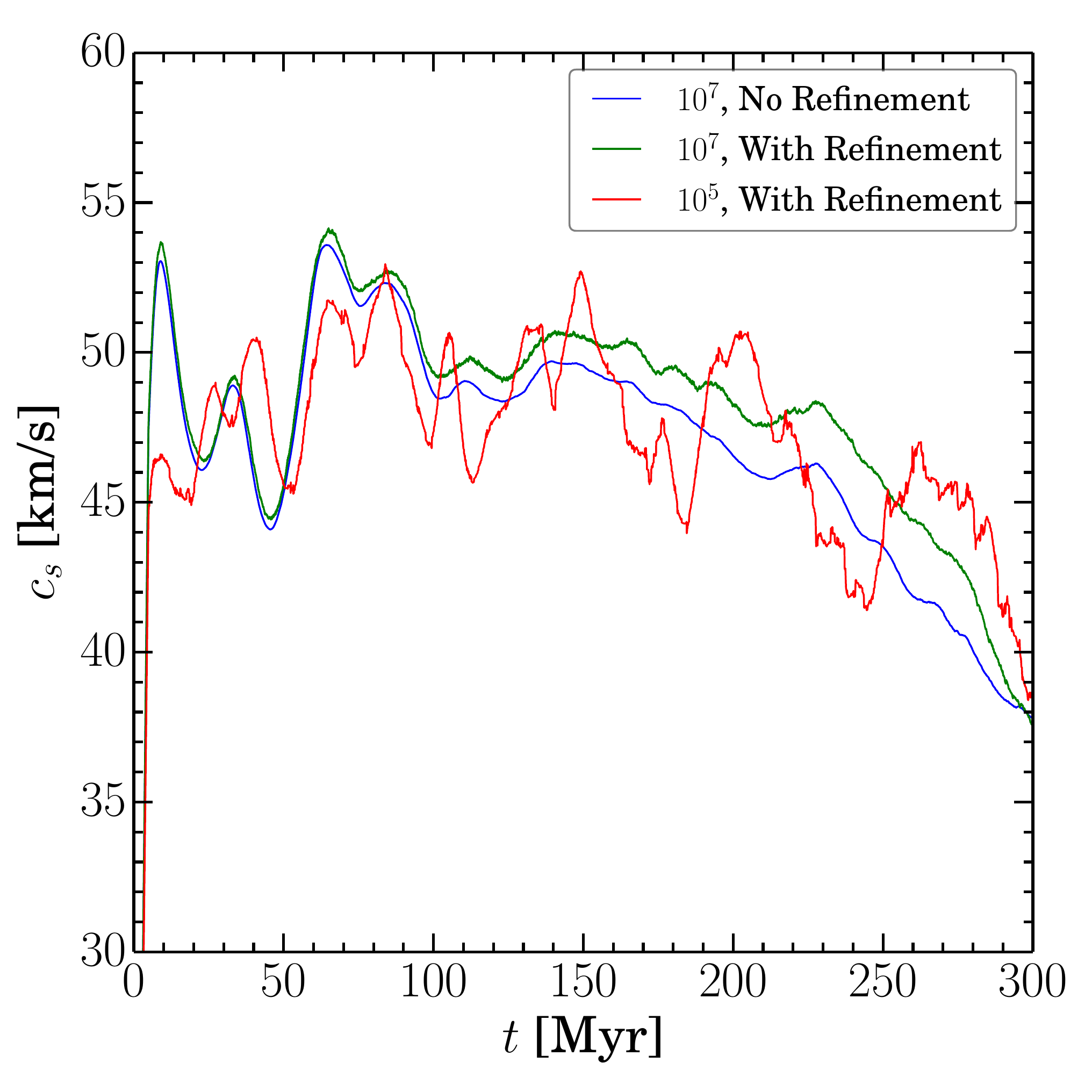}
\includegraphics[width=0.32\textwidth]{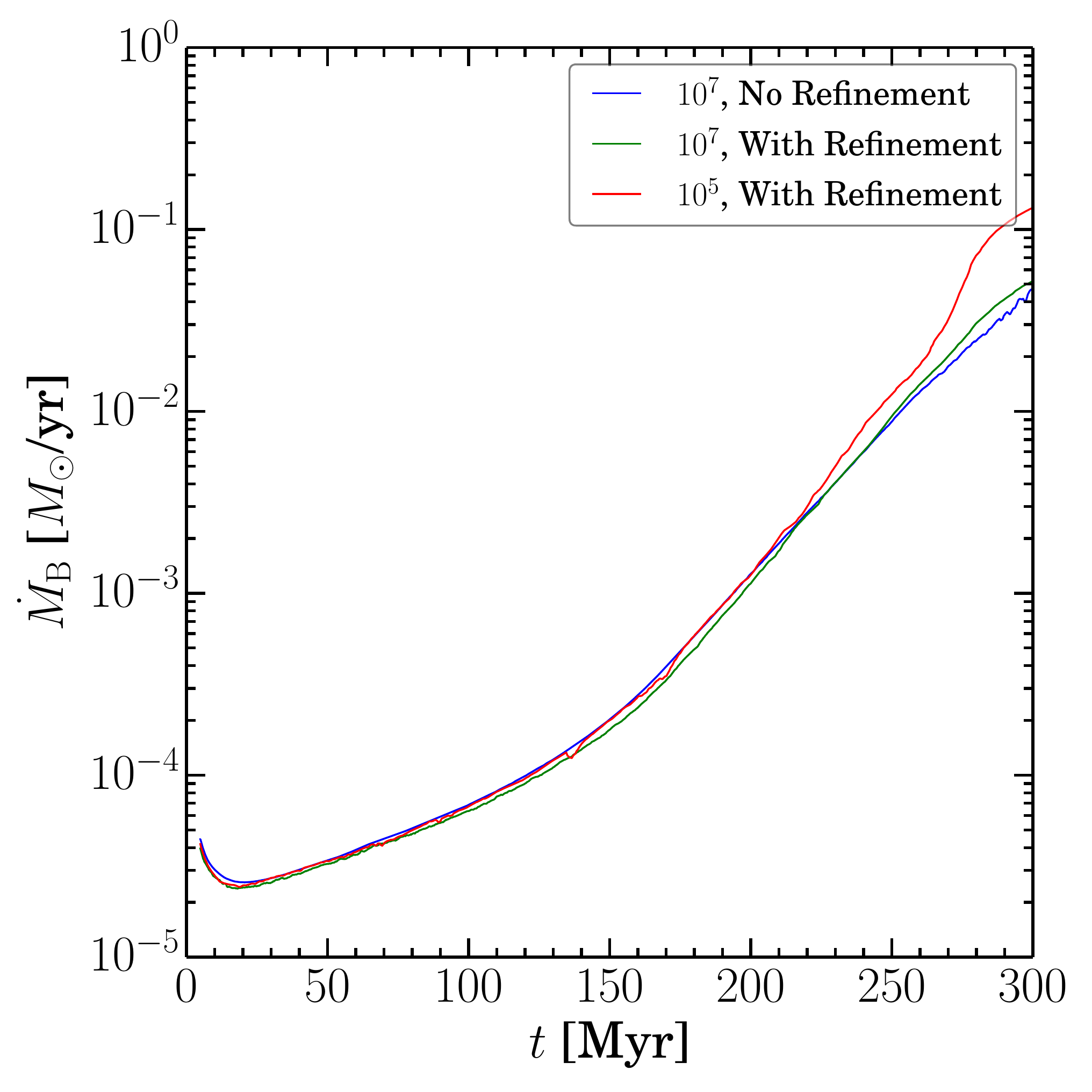}
  \caption{Convergence of refinement scheme. Here, we show the fluid
    parameters (estimated at the full smoothing length of the black hole) and
    Bondi-Hoyle rate in simulations of an isolated disc galaxy 
    with no black hole feedback. We plot two high resolution runs using $10^7$
    resolution elements, with and without refinement, that show good agreement
    with each other, and also with the lower resolution run ($10^5$ resolution
    elements) that includes black hole refinement.}
 \label{hsml_high_comp}
\end{figure*}

\end{document}